\documentclass[aps,pre,twocolumn,superscriptaddress,aps]{revtex4}
\usepackage[dvips]{graphics}
\usepackage{graphicx}
\usepackage{amsfonts}
\usepackage{amssymb}
\usepackage{amsmath}
\usepackage{subfigure}
\usepackage{color}
\newcommand{\cP}{\ensuremath{\mathcal{P}}}
\newcommand{\cT}{\ensuremath{\mathcal{T}}}

\begin{document}

\title{$\cP\cT$-symmetric double-well potentials revisited: bifurcations,
stability and dynamics}

\author{A.\ S.\ Rodrigues}
\affiliation{Departamento de F\'{\i}sica/CFP, Faculdade de Ci\^{e}ncias,
Universidade do Porto, R. Campo Alegre, 687 - 4169-007 Porto, Portugal}

\author{K. Li}
\affiliation{Department of of Mathematics and Statistics, University of
Massachusetts, Amherst, MA 01003-9305, USA}

\author{V. Achilleos}
\affiliation{Department of Physics, University of Athens, Panepistimiopolis,
Zografos, Athens 157 84, Greece}

\author{P. G. Kevrekidis}
\affiliation{Department of of Mathematics and Statistics, University of
Massachusetts, Amherst, MA 01003-9305, USA}

\author{D.\ J.\ Frantzeskakis}
\affiliation{Department of Physics, University of Athens, Panepistimiopolis,
Zografos, Athens 157 84, Greece}

\author{Carl M.\ Bender}
\affiliation{Department of Physics, Washington University, St. Louis, MO 63130,
USA}

\begin{abstract}
In this work we analyze $\cP\cT$-symmetric double-well potentials based on a
two-mode picture. We reduce the problem into a $\cP\cT$-symmetric dimer and
illustrate that the latter has effectively two fundamental bifurcations, a
pitchfork (symmetry-breaking bifurcation) and a saddle-center one, which is the
nonlinear analog of the $\cP\cT$-phase-transition. It is shown that the symmetry
breaking leads to {\it ghost states} (amounting to growth or decay); although
these states are not true solutions of the original continuum problem, the
system's dynamics closely follows them, at least in its metastable evolution.
Past the second bifurcation, there are no longer states of the original
continuum system. Nevertheless, the solutions can be analytically continued to
yield a new pair of branches, which is also identified and dynamically examined.
Our explicit analytical results for the dimer are directly compared to the full
continuum problem, yielding a good agreement.
\end{abstract}
\maketitle

\section{Introduction}
$\cP\cT$-symmetric quantum systems~\cite{R1,R2} have emerged as an intriguing
complex generalization of conventional quantum mechanics and have been a focus
for numerous investigations at the interface between theoretical physics and
applied mathematics. The key premise is that fundamental physical symmetries
such as parity $\cP$ and time reversal $\cT$ may be sufficient (in suitable
parametric regimes) to ensure that the eigenvalues of the Hamiltonian are real.
Thus, $\cP\cT$-symmetric Hamiltonians provide an alternative to the standard
postulate that the Hamiltonian operator be Dirac Hermitian (invariant under
matrix transposition and complex conjugation $*$). In the context of
Schr\"odinger Hamiltonians with a complex potential $V(x)$, the constraint of
$\cP\cT$ symmetry requires that the potential satisfy $V(x)=V^*(-x)$; that is,
$V(x)$ has a symmetric real part and an antisymmetric imaginary part.

On the other hand, {\it nonlinear} Schr\"odinger (NLS) equations incorporating
double-well potentials have received attention due to applications in atomic and
optical physics. Such potentials can easily be realized in the context of atomic
Bose-Einstein condensates (BECs) through the combination of a parabolic
(harmonic) trap with a periodic potential. Experiments have observed fundamental
effects, including tunneling and Josephson oscillations for small atom numbers,
macroscopic quantum self-trapped states for large atom numbers \cite{R3}, and
nonlinearity-induced symmetry-breaking dynamical instabilities \cite{R4}.
Theoretical studies accompanying these developments have examined finite-mode
reductions, analysis of the bifurcations and their dynamical implications
\cite{R5,R6,R7,R8,R9,R10,R11,R12}, as well as quantum effects \cite{R13} and
nonlinear variants of the potentials \cite{R14}. A similar phenomenology has
also been found in nonlinear optical settings, with results for the formation of
asymmetric states in dual-core fibers \cite{R15}, self-guided laser beams in
Kerr media \cite{R16}, and optically-induced dual-core waveguide structures in
photorefractive crystals \cite{R17}.

Recently, double-well potentials in the context of $\cP\cT$-symmetric nonlinear
systems have received considerable attention. This is due to the pioneering work
of Christodoulides and collaborators, who proposed that nonlinear optics
presents a fertile ground for the experimental realization of $\cP\cT$-symmetric
systems. The first realization of $\cP\cT$-symmetry in a waveguide coupler arose
in the so-called passive-$\cP\cT$ setting in which two waveguides, one with loss
and the other without loss, were used \cite{R18}. A similar proposal for the
existence of a leaking dimer (in the presence of nonlinearity) was formulated in
the atomic setting of Bose-Hubbard models \cite{R19}. Subsequently, an optical
waveguide system with both gain and loss was studied and the role of the
nonlinearity in the dynamics was explored \cite{R20}. Further experimental
investigations were concerned with electrical analogs of the system \cite{R21}.
Theoretical investigations have rapidly followed by examining such dimer-type
settings \cite{R22,R23,R24,R25,R26,R27,R28,R29,R30} and generalizations thereof,
including ones where the gain-loss contributions appear in a balanced form in
front of the nonlinear term \cite{R31,R32,R33}.

This paper revisits this theme of $\cP\cT$-symmetric double-well potentials. Our
motivation is to unify the above studies with an important recent contribution,
namely, Ref.~\cite{R34}. To be specific, in the early works on NLS models with
double-well potentials \cite{R35,R5,R11,R12}, and also in the $\cP\cT$-symmetric
dimer \cite{R28}, the symmetry-breaking bifurcation was identified, but the
asymmetric states that normally accompany the bifurcation \cite{R5,R11,R12}
could not be identified. The simpler system of a $\cP\cT$-symmetric pair of
$\delta$-function potentials, where the solution can be obtained by means of a
five-dimensional numerical root search, was studied in Ref.~\cite{R34}. There,
it was found that the bifurcation results in the emergence of what we call a
{\it ghost state}, namely, a solution of the steady-state problem with a {\it
complex} nonlinear eigenvalue parameter (complex propagation constant in optics
or complex chemical potential in BECs). [These states were discussed in a 2006
paper for the case of the (discrete) leaking dimer \cite{R19}.] Another
observation of Ref.~\cite{R34} concerned the possibility of performing an {\it
analytic continuation} of the symmetric and antisymmetric solutions of the
original double-well problem {\it past the point of their $\cP\cT$ phase
transition}; that is, past the critical point of the saddle-center bifurcation
where they disappear simultaneously. This is in the broader spirit of $\cP
\cT$-symmetric quantum mechanics \cite{R2}.

In this paper we study an NLS model with a double-well potential, which has an
even real part and a $\cP\cT$-symmetric imaginary part. We combine the above
ideas, interweaving analytics based on the two-mode analysis of the simple dimer
model and numerics in the framework of the partial differential equation (PDE).

This paper is organized as follows. We present the two-mode analysis of the
discrete nonlinear Schr\"odinger equation (DNLS) dimer in Sec.~II, where we
examine the symmetric/antisymmetric and asymmetric states and the potential
for nonlinearity-induced symmetry-breaking bifurcations. We show how the $\cP
\cT$-symmetric variant of the dimer problem adjusts these features. The modified
critical point of the pitchfork bifurcation is identified, and the feature of a
saddle-center bifurcation, which is a nonlinear analog of the $\cP\cT$ phase
transition, is discussed in the spirit of some of the earlier work on this
theme. The key by-products of these bifurcations, namely, the ghost states
emanating from the pitchfork and the analytically continued states past the
nonlinear analog of the $\cP\cT$ phase transition, are found in {\it analytical}
form. Armed with these analytical results, we study the full PDE problem in
Sec.~III, where the aim is to reduce the PDE, through suitable and fully
quantified approximations and transformations (so that one can translate the
dimer results to the PDE and vice-versa), to a $\cP\cT$-symmetric dimer. In
Sec.~IV we compare the analytical results concerning the dimer problem with
numerical ones obtained in the framework of the original NLS model. In addition
to quantifying the bifurcations and identifying their ``daughter'' states, we
examine the dynamics of the various unstable states within the system. Finally,
in Sec.~V we briefly summarize our findings and present some possibilities for
further study.

\section{DNLS dimer}

\subsection{Classical DNLS dimer}

We begin our exposition by revisiting a simpler initial problem that has been
long studied (see, e.g., Ref.~\cite{R35}), namely the DNLS dimer of the form
\begin{eqnarray}
i\dot{u}_1&=&-ku_2-|u_1|^2u_1,\nonumber \\ i\dot{u}_2&=&-ku_1-|u_2|^2u_2,
\label{e1}
\end{eqnarray}
where overdots denote derivatives with respect to the evolution variable $t$ 
(which denotes the propagation distance in optics). As is customary, we seek
stationary solutions of the form $u_1=\exp(iEt)a$ and $u_2=\exp(iEt)b$, where
the complex amplitudes $a$ and $b$ and the propagation constant $E$ are
determined by the equations
\begin{equation}
Ea=kb+|a|^2a,\qquad Eb=ka+|b|^2b.
\label{e2}
\end{equation}
Using a polar representation of the two ``sites'', namely $a=Ae^{i\theta_a}$,
$b=Be^{i\theta_b}$, we immediately obtain $\sin\theta=0$, where $\theta=
\theta_b-\theta_a$, and hence the resulting equations for the amplitudes are
\begin{equation}
EA=\pm kB+A^3,\qquad EB=\pm kA+B^3.
\label{e3}
\end{equation}
The simpler solutions of the above system are symmetric ones of amplitudes $A^2=
B^2=E\mp k$ with the upper (lower) sign corresponding to the in-phase
(out-of-phase) profile. In addition, there exist asymmetric solutions of
amplitudes $A\neq B$, which can be determined by the following equations
derived by eliminating $E$ from (\ref{e3}):
\begin{equation}
AB=\pm k,\qquad A^2+B^2=E.
\label{e4}
\end{equation}
The amplitudes $A^2=(E\pm\sqrt{E^2-4k^2})/2$, $B=k/A$ exist {\it only} for $E^2>
4k^2$ and thus for $E>2k$ or $E<-2k$. These asymmetric solutions coincide with
the symmetric ones at $E^2=4k^2$, and hence a {\it pitchfork} symmetry-breaking
bifurcation is responsible for their emergence in the nonlinear problem. As is
known from previous works~\cite{R5,R11,R12}, this bifurcation arises for the
focusing (attractive) nonlinearity case $k>0$ at the point $E=2k$, and does so
from the symmetric branch. However, in the defocusing (repulsive) case $k<0$ the
bifurcation arises when $E=-2k$ and the asymmetric solution emerges from the
antisymmetric branch. This bifurcation picture is also complemented by the
stability eigenvalues for the focusing and defocusing cases below (see
Sec.~II.B).

\subsection{$\cP\cT$-symmetric DNLS dimer}

We now turn to the $\cP\cT$-symmetric variant of the DNLS dimer, as it was
recently analyzed in Ref.~\cite{R28} (see also Ref.~\cite{R22} for the
experimental investigation of Ref.~\cite{R20}). This setting is described by the
system
\begin{eqnarray}
i\dot{u}_1&=&-ku_2-|u_1|^2u_1-i\gamma u_1,\nonumber \\
i\dot{u}_2&=&-ku_1-|u_2|^2u_2+i\gamma u_2,
\label{e5}
\end{eqnarray}
which incorporates matched linear loss and gain of strength $\gamma$ acting on
the components $u_1$ and $u_2$. The stationary equations have the form
\begin{eqnarray}
E a&=&k b+|a|^2a+i\gamma a,\nonumber\\
E b&=&k a+|b|^2b-i\gamma b.
\label{e6}
\end{eqnarray}

Using the same polar decomposition as before, we find that the amplitudes and
phase difference of the symmetric solutions satisfy
\begin{eqnarray}
A^2&=&B^2=E \pm \sqrt{k^2-\gamma^2}, \label{e7} \\
\sin\theta&=&-\gamma/k, \label{e8}
\end{eqnarray}
where the signs in (\ref{e7}) correspond to the first ($+$) and second ($-$)
branches. These two solutions exist {\it only} up to $\gamma=k$, while there is
no such limit in the Hermitian case $\gamma=0$. The latter is the critical
threshold for the $\cP\cT$ symmetry breaking of the underlying linear problem,
whose eigenvalues are $E=\pm\sqrt{k^2 -\gamma^2}$. Beyond this critical value
the eigenvalues become imaginary. Additionally, as is shown in Ref.~\cite{R28},
the stability eigenvalues of this $\cP\cT$-symmetric dimer are
$$\pm2i\left[2(k^2-\gamma^2)-E\sqrt{k^2-\gamma^2}\right]^{1/2}$$
for the $(-)$ branch and
$$\pm2i\left[2(k^2-\gamma^2)+E\sqrt{k^2-\gamma^2}\right]^{1/2}$$
for the $(+)$ branch.

For the case $\gamma=0$ (see Sec.~II.A), these eigenvalues describe the critical
value of $E$, where the two branches become unstable, and the pitchfork
bifurcation leading to the asymmetric states emerges. Specifically, if $k>0$
(focusing nonlinearity), the first branch corresponding to the symmetric
solutions becomes unstable at $E=2k$; if $k<0$ (defocusing nonlinearity), the
destabilization arises when $E=-2k$, and this happens for the second branch
corresponding to the antisymmetric branch $\theta=\pi$.

Similar stability conclusions occur for $\gamma\ne0$, where only the $(-)$
branch becomes unstable, but now also for $\gamma^2 \geq k^2- E^2/4$. This
suggests that there are two possibilities. If the propagation constant $E$ (and
the coupling strength $k$) are such that $E^2 < k^2/4$, then the instability is
induced by the increase of the $\cP\cT$-symmetry parameter $\gamma$ at the
critical point. However, if $E^2>4k^2$, the instability has ``already'' taken
place due to the presence of nonlinearity, and the $(-)$ branch is unstable even
in the $\gamma=0$ limit. In the latter case the presence of the gain-loss aspect
only enhances the instability.

What has become of the pitchfork bifurcation picture explored earlier? We can
see that the same instability is present here (at least if $E^2>4k^2$). However,
analogs of the symmetry-broken states past the critical point, i.e., stationary
asymmetric states (emerging after the instability of the symmetric states),
cannot be identified. Inevitably, the question of their fate arises. This type
of question was initially raised in Ref.~\cite{R19} (where a leaky quantum
dimer, with loss only, was considered) and, past a critical point, states with a
complex (instead of real) ``eigenvalue'' $E$ were identified. In the $\cP
\cT$-symmetric context, a similar idea was put forth in Ref.~\cite{R34} for a
double well consisting of two delta functions. In our paper we unify the
approaches of Refs.~\cite{R19,R34} by computing the ``ghost states'' (as we
characterize them) that emerge from the symmetry-breaking bifurcation.

Before presenting the computation of the ghost states, we comment that these
states with complex (nonlinear) eigenvalue $E$ are no longer {\it true}
solutions of the original system (\ref{e5}). This is because of the U$(1)$
invariance of the system, which only allows stationary solutions with real $E$
[so that $|\exp(iEt)|^2=1$]. When this symmetry is violated, the solutions may
satisfy the stationary equations (\ref{e6}), but are only ghost states of the
original dynamical system because they do not satisfy (\ref{e5}). Thus, at best
one expects that the dynamics may stay close to the dynamics of these ghost
states, especially during the evolution of the symmetry-breaking instability. We
return to this topic later.

To identify these stationary solutions, we introduce polar coordinates $E\to
E\exp(i\phi_E)$ and get
\begin{eqnarray}
EA\cos(\phi_E) &=& kB\cos\theta+A^3,\nonumber\\
EB\cos(\phi_E) &=& kA\cos\theta+B^3, \label{e9}\\
E\sin(\phi_E) &=& k\frac{B}{A}\sin\theta+\gamma,\nonumber\\
E\cos(\phi_E) &=& -k\frac{A}{B}\cos\theta-\gamma. \label{e10}
\end{eqnarray}
To derive the asymmetric ($A\neq B$) solutions, we rewrite these equations as
\begin{eqnarray}
\cos\theta &=& AB/k,\nonumber\\
\sin\theta &=& -\frac{2 \gamma}{k}\frac{A B}{A^2+B^2}, \label{e11} \\
E\cos(\phi_E) &=& A^2+B^2,\nonumber\\
E \sin(\phi_E) &=& \gamma\frac{A^2-B^2}{A^2+B^2}. \label{e12}
\end{eqnarray}
Applying the identity $\sin^2\chi+\cos^2\chi=1$ to (\ref{e11}), we obtain a
condition for the solution amplitudes, and the same identity applied to
(\ref{e12}) yields the parameter $E$. These solutions exist {\it only} for
$\gamma^2>k^2-E^2/4$. If $E^2>4k^2$, they exist for all values of $\gamma$
(i.e., they have bifurcated ``already'' due to the nonlinearity). Also, these
solutions terminate as $\theta\to-\pi/2$, $\phi_E\to\pi/2$. In turn, this
implies that in this limit both $B$ and $A$ vanish, with the ratio between them
having the limit $B/A\to(\gamma\pm\sqrt{\gamma^2-k^2})/k$. Thus, we have
identified the disappearance point $\gamma^2=E^2+k^2$ of these symmetry-broken
solutions.

Finally, we discuss the disappearance of the two symmetric states at the
critical point $\gamma=k$, which is the phase-transition point of the linear
(and nonlinear) problem. We have shown that at this point the symmetry-broken
states still exist, but that now they are only ghost states of the steady-state
problem. From the point of view of nonlinear theory, one may be content to find
a saddle-center bifurcation at this point, which leads to the disappearance of
these solutions as stationary states of the nonlinear problem. Yet, once again,
when these solutions disappear (even in the normal form of such a bifurcation),
this means that they appear somewhere else within the complex plane of
solutions. In order to compare this result with the linear $\cP\cT$-symmetric
case, where the eigenvalues collide, become complex, and continue to exist in
the complex plane, we follow Ref.~\cite{R34} and consider the analytic
continuation of our solutions. In the $\cP\cT$-symmetric regime up to the
critical point, the solutions are chosen so that $u^*(x)=u(-x)$. (This is a
broader statement for a spatially distributed system; in our simpler dimer
setting, we need only replace $x$ by the subscript $1$ and $-x$ by the subscript
$2$, or vice-versa.) Thus, to perform the analytic continuation, we use $u_j^*=
u_{3-j}$ in (\ref{e5}), which leads to (\ref{e9})-(\ref{e10}), but in the first
pair of equations $A^3$ and $B^3$ are replaced by $A^2B$ and $B^2A$. The result
is
\begin{eqnarray}
\xi &=& \frac{B}{A}=\frac{\gamma \pm \sqrt{\gamma^2-k^2}}{k},\nonumber\\
A^4 &=& \frac{E^2-\gamma^2\left(\frac{1-\xi^2}{1+\xi^2}\right)^2}{\xi^2},
\label{e13} \\
E\sin(\phi_E)&=&\gamma\frac{1-\xi^2}{1+\xi^2}=\pm\sqrt{\gamma^2-k^2},\nonumber\\
\theta&=&-\pi/2. \label{e14}
\end{eqnarray}
Note that the pitchfork symmetry-breaking branches also tend to this solution,
as shown above in the expression for $B/A$ in the limit of termination of the
branch when $\gamma^2=E^2+k^2$. We have made an additional subtle assumption
here, namely, that $\theta_a+\theta_b=0$. We can obtain more general solutions
without this assumption, but these do not appear
to introduce new features to the problem.

The solutions stemming from the analytic continuation provide a complete
description of the states of the system. We move from symmetric/antisymmetric
states to asymmetric ones (which may be ghost states) through a pitchfork
bifurcation, destabilizing the symmetric (antisymmetric) branch for a focusing
(defocusing) nonlinearity. We terminate at the point of the linear $\cP\cT$
phase transition, where the nonlinear eigenvalues of the symmetric branches
collide and become complex, giving rise to an analytic continuation of our
solutions in the complex plane. All solutions terminate at $\gamma^2=k^2+E^2$.

\section{From the NLS equation with a double-well potential to the dimer}

The central problem of interest here is an NLS equation with a double-well
potential,
\begin{eqnarray}
iu_t={\cal L}u+iV_{\cP\cT}u+|u|^2u,
\label{e15}
\end{eqnarray}
where $u_t=\partial u/\partial t$. Here, $u(x,t)$ is a complex field (which can
represent the electric field envelope in optics or the macroscopic wavefunction
in BECs); ${\cal L}=-(1/2)\partial_x^2+V_{\rm real}(x)$ is a linear
Schr\"odinger operator containing a real, symmetric double-well potential $V_{
\rm real}(x)$ (see, e.g., Refs.~\cite{R5,R12}); $iV_{\cP\cT}$ is a purely
imaginary odd potential with $V_{\cP\cT}(-x)=-V_{\cP\cT}(x)$.

\subsection{Two-mode reduction}

In such settings two-mode approximations have been valuable tools for studying
the statics, stability, and dynamics of the system (see rigorous justifications
in Refs.~\cite{R36,R37,R38}). However, we do not use a Galerkin truncation to
the ground and excited eigenmodes $\{u_0,u_1\}$ of the linear operator ${\cal
L}$ with eigenvalues $\omega_0$, $\omega_1$ \cite{R12,R11,R36,R37,R38}, but
rather the rotated basis $\{u_L,u_R\}$ \cite{R5,R39}
\begin{equation}
u_L=\left(u_0-u_1\right)/\sqrt{2},\qquad u_R=\left(u_0+u_1\right)/\sqrt{2}.
\label{e16}
\end{equation}
The subscripts $L$ and $R$ denote the left and right well of $V_{\rm real}$. 
Note that $u_0$ and $u_1$ (the ground- and first-excited state) are even and
odd functions of $x$.

Following Ref.~\cite{R39}, we approximate the solution of (\ref{e15}) by the
Galerkin expansion $u(x,t)=c_L(t)u_L(x)+c_R(t)u_R(x)$, where $c_{L,R}$ are
unknown time-dependent coefficients. Substituting this ansatz into (\ref{e15})
and subsequently projecting on $u_L$, $u_R$, we obtain the following equations
for $c_{L,R}$:
\begin{eqnarray}
i\dot{c}_L &=& \Omega c_L-\omega c_R+i\gamma_Lc_L+\eta_L|c_L|^2c_L,\nonumber\\
i\dot{c}_R &=& \Omega c_R-\omega c_L+i\gamma_Rc_R+\eta_R|c_R|^2c_R,\label{e17}
\end{eqnarray}
where $\Omega=(\omega_0+\omega_1)/2$ and $\omega=(\omega_1-\omega_0)/2$. Also,
$\gamma_{L,R}=\int dx\,V_{\cP\cT}(x)u_{L,R}^2$, and due to the parities of
$u_{0,1}$ and $V_{\cP\cT}$ it follows that $\gamma_L=-\gamma_R\equiv\gamma$. To
derive the system above we have assumed, in addition to the truncation itself,
that the overlap integrals $\int dx\,\phi_L^2\phi_R^2$, $\int dx\,\phi_L^3\phi_R
$, and $\int dx\,\phi_R^3\phi_L$ are negligible in comparison to $\eta_L\equiv
\int dx\,\phi_L^4$ and $\eta_R\equiv\int dx\,\phi_R^4$. This approximation
becomes better as the distance between the wells becomes larger because these
three integrals depend exponentially on the separation between the wells because
of the exponential decay of the bound states $u_{L,R}(x)$. (Comparisons of these
terms with the dominant terms $\eta_{L}$, $\eta_R$ can be found in
Ref.~\cite{R39}, which attests to the validity of this approximation.) Note
that, as shown in Ref.~\cite{R12}, even the full model with these additional
terms is analytically tractable. However, for our present purposes, this
assumption considerably simplifies the analysis and helps to connect with the
study of the $\cP\cT$-symmetric DNLS dimer. The equality $\eta_L=\eta_R\equiv
\eta$ mirrors the symmetric nature of $V_{\rm real}(x)$, while $\gamma_L=-
\gamma_R\equiv\gamma$, mirrors the antisymmetric nature of $V_{\cP\cT}(x)$.

We now consider solutions to (\ref{e17}) of the form
\begin{equation}
c_{L,R}(t)=\eta^{-1/2}C_{L,R}\exp(-i\mu t), 
\label{e18}
\end{equation}
where $C_{L,R}$ are unknown constant amplitudes and $\mu$ is the propagation
constant (chemical potential in BECs). The prefactor $\eta^{-1/2}$ is used to
rescale the term in front of the nonlinearity. Substituting this ansatz into
(\ref{e17}) and using $E=\mu-\Omega$ as the (nonlinear) eigenvalue, we find that
$C_{L,R}$ satisfy precisely the stationary equations for the DNLS $\cP
\cT$-symmetric dimer
\begin{eqnarray}
EC_L&=& kC_R+|C_L|^2C_L+i\gamma C_L,\nonumber\\
EC_R&=& kC_L+|C_R|^2C_R-i\gamma C_R,
\label{e19}
\end{eqnarray}
where we have used the notation $-\omega=k$.

To further justify our assumptions regarding the overlap integrals, we note that
for the Hamiltonian case of $\gamma=0$, the symmetry breaking is predicted to
occur at $E=\mu-\Omega=2k=-2\omega$ for the focusing nonlinearity case and at
$\mu-\Omega=-2k=2 \omega$ for the defocusing one. This implies a bifurcation at
$\mu=(3\omega_0-\omega_1)/2$ in the focusing case and at $\mu=(3\omega_1-
\omega_0)/2$ in the defocusing case. The corresponding predictions for the
bifurcations induced by nonlinearity for (\ref{e15}) are $\mu=\omega_0-A_0(
\omega_1-\omega_0)/(3B-A_0)$ for the focusing case, while the critical $\mu$ for
the defocusing one is $\mu=\omega_0+3B(\omega_1-\omega_0)/(3B-A_1)$ \cite{R12}.
In these expressions $A_0=\int dx\,u_0^4$, $A_1=\int dx\,u_1^4$, while $B=\int
dx\,u_0^2u_1^2$. Hence, it is clear that the approximation for the overlap
integrals made above is tantamount to $A_0=A_1=B$ because in that case the two
expressions (the one from the dimer and the one from the exact two-mode
reduction) coincide, showing that this is a reasonable approximation and that it
trades a simple prediction for the critical points, which can be quantified
just by knowing the eigenvalues of the underlying linear problem, for a small
loss of accuracy in the result. This is the approximation that we use for the
full problem.

\subsection{Direct simulations versus analytical approximations}

Having reduced (\ref{e15}) to the simple dimer model of (\ref{e19}), we now use
the results of Sec.~II to study the bifurcations of our $\cP\cT$-symmetric NLS
model. To make connections between our work with earlier papers on double-well
potentials, we use the same form of the double-well potential that was used in
Refs.~\cite{R12,R39}. It should, however, be evident from the exposition above
that our basic phenomenology and corresponding conclusions will be quite
general. In particular, we consider a potential of the form $V(x)=V_{\rm real}(x
)+iV_{\cP\cT}(x)$, where
\begin{eqnarray}
V_{\rm real}(x)&=& \frac{1}{2}\Omega_{\rm tr}^2 x^2+V_0{\rm sech}^2
\left(\frac{x}{w}\right), \nonumber\\
V_{\cP\cT}(x) &=& \varepsilon x\exp\left(-\frac{x^2}{2}\right).
\label{e20}
\end{eqnarray}
The real (even) potential $V_{\rm real}(x)$ consists of a parabolic trap of
strength $\Omega_{\rm tr}$ and a localized barrier potential of strength $V_0$
and width $w$, which has a standard double-well structure (that is easily
achievable, e.g., in atomic BECs). The imaginary potential $V_{\cP\cT}(x)$ of
strength $\varepsilon$ has a prototypical antisymmetric gain-loss profile
\cite{R40}. Hereafter, we fix the parameters of $V_{\rm real}$ to be $\Omega_{
\rm tr}=0.1$, $V_0=1$ and $w=0.5$, which yields the two lowest eigenvalues of
the potential as $\omega_0=0.13282$ and $\omega_1=0.15571$. This imposes a
tunneling strength between the wells of $k=-0.01145$.

There are now two possible paths to follow, both of which yield the
symmetry-breaking pitchfork bifurcation, as indicated above. The first involves
increasing the propagation constant $\mu$ (and hence the dimer eigenvalue $E$),
thus increasing the strength of the nonlinearity. The second involves increasing
the $\cP\cT$-symmetric potential parameter $\varepsilon$ (and hence the
corresponding dimer parameter $\gamma$). Both of these paths will yield the
pitchfork bifurcation when $\gamma^2=k^2-E^2/4$ if the nonlinearity is
sufficiently weak (namely, $E^2<4k^2$). However, only the latter path, which we
follow below, exhibits the $\cP\cT$-symmetry-breaking phase transition when
$\gamma=\pm k$.

\begin{figure}[tbp]
\includegraphics[width=78mm,keepaspectratio]{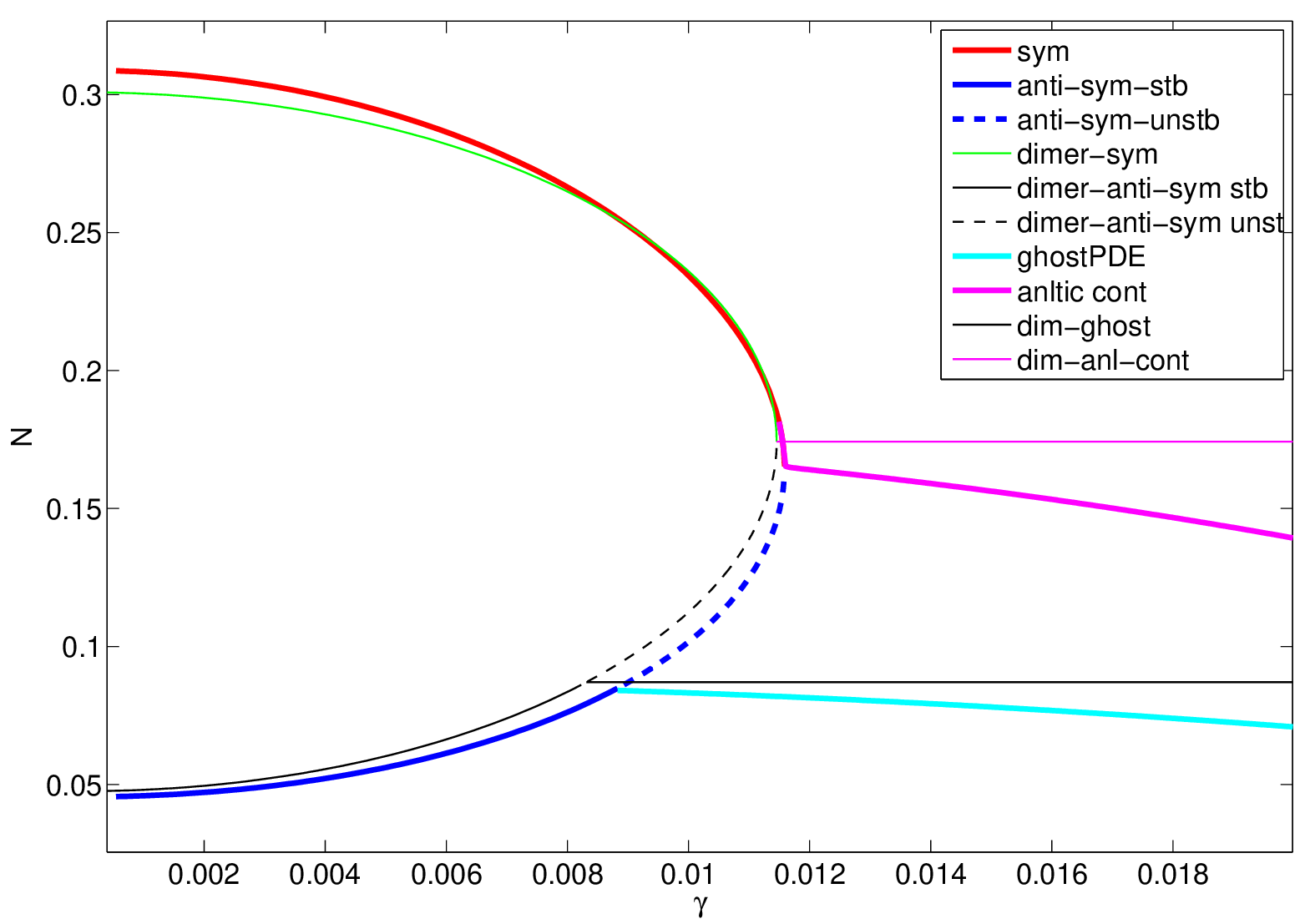}
\includegraphics[width=78mm,keepaspectratio]{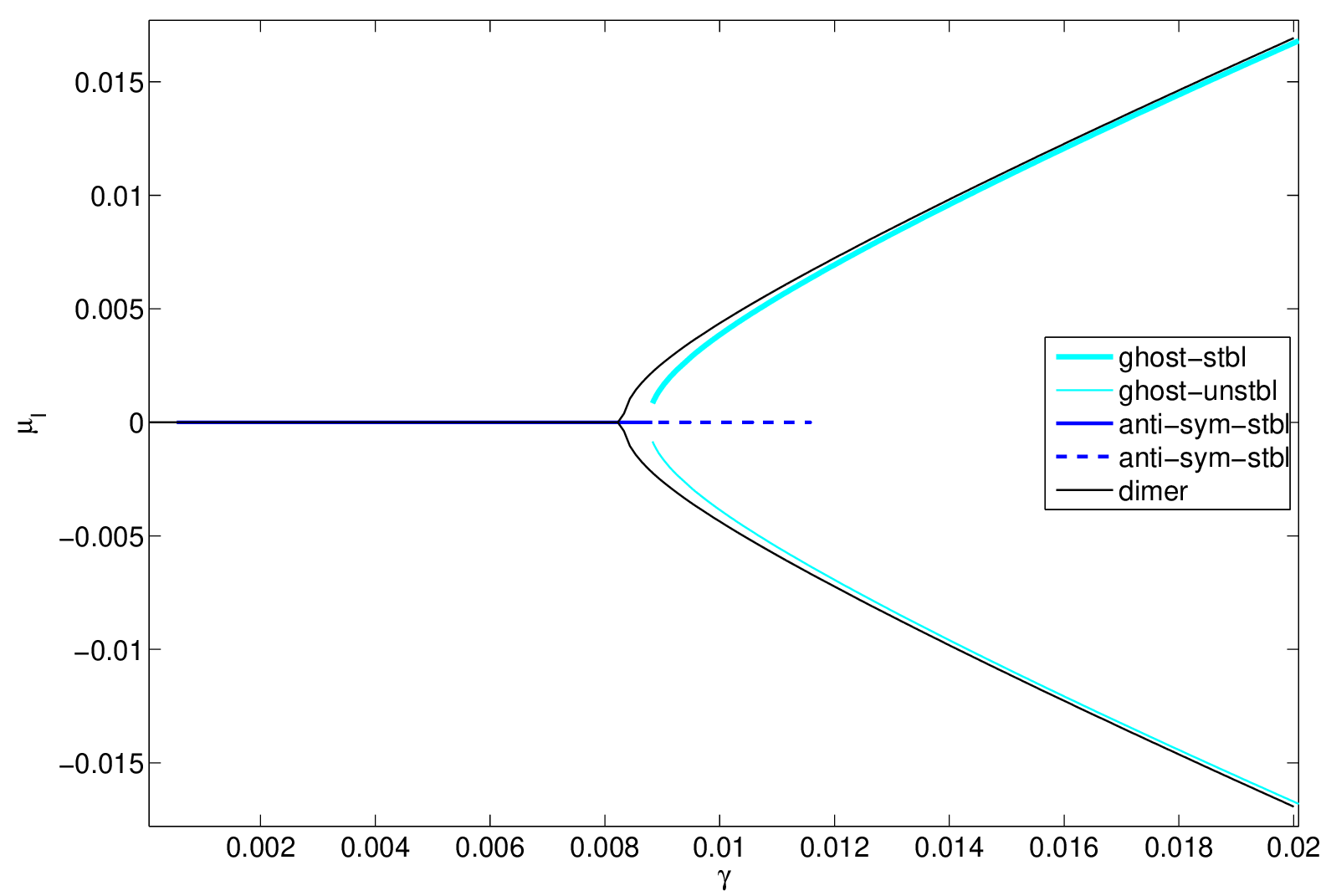}
\caption{(Color online)
Top panel: Bifurcation diagram $N=N(\gamma)$. The thin lines are the predictions
of the two-mode analysis (the solid line denotes a linearly stable branch, while
the dashed line depicts a linearly unstable one). The thick lines represent the
full PDE analog (with the same stability designation). Bottom panel: Bifurcation
diagram $\mu_I=\mu_I(\gamma)$. The emergence of the ghost branches is shown as a
supercritical pitchfork bifurcation in these variables. Once again, a comparison
to the two-mode picture is given; $\mu_I$ is computed according to the last
equation of (\ref{e12}), leading to $\mu_I=\pm\gamma\sqrt{(\mu-\Omega)^2+4
\gamma^2-4k^2}/\sqrt{(\mu-\Omega)^2+4\gamma^2}$.}
\label{fig1}
\end{figure}

In Fig.~\ref{fig1} we show the bifurcation diagram $N=N(\gamma)$, where $N=
\int_{-\infty}^\infty dx\,|u|^2$ represents the energy in optics (number of
atoms in BECs). The figure compares the full PDE model with the $\cP
\cT$-symmetric dimer model, where the real part of the chemical potential is
chosen to be $\mu=0.16$. This choice of $\mu$ is justified by the requirement to
be sufficiently close to the linear eigenvalues so that the two-mode picture is
valid. Given the defocusing nature of the nonlinearity used, $\mu>\omega_1$. A
focusing variant of the problem was studied in Ref.~\cite{R34}; based on our
comments in the previous sections, it is easy to adapt the results below (upon
the exchange of the antisymmetric branch with the symmetric branch as the parent
branch in the pitchfork bifurcation).

\begin{figure}[tbp]
\includegraphics[width=41mm,keepaspectratio]{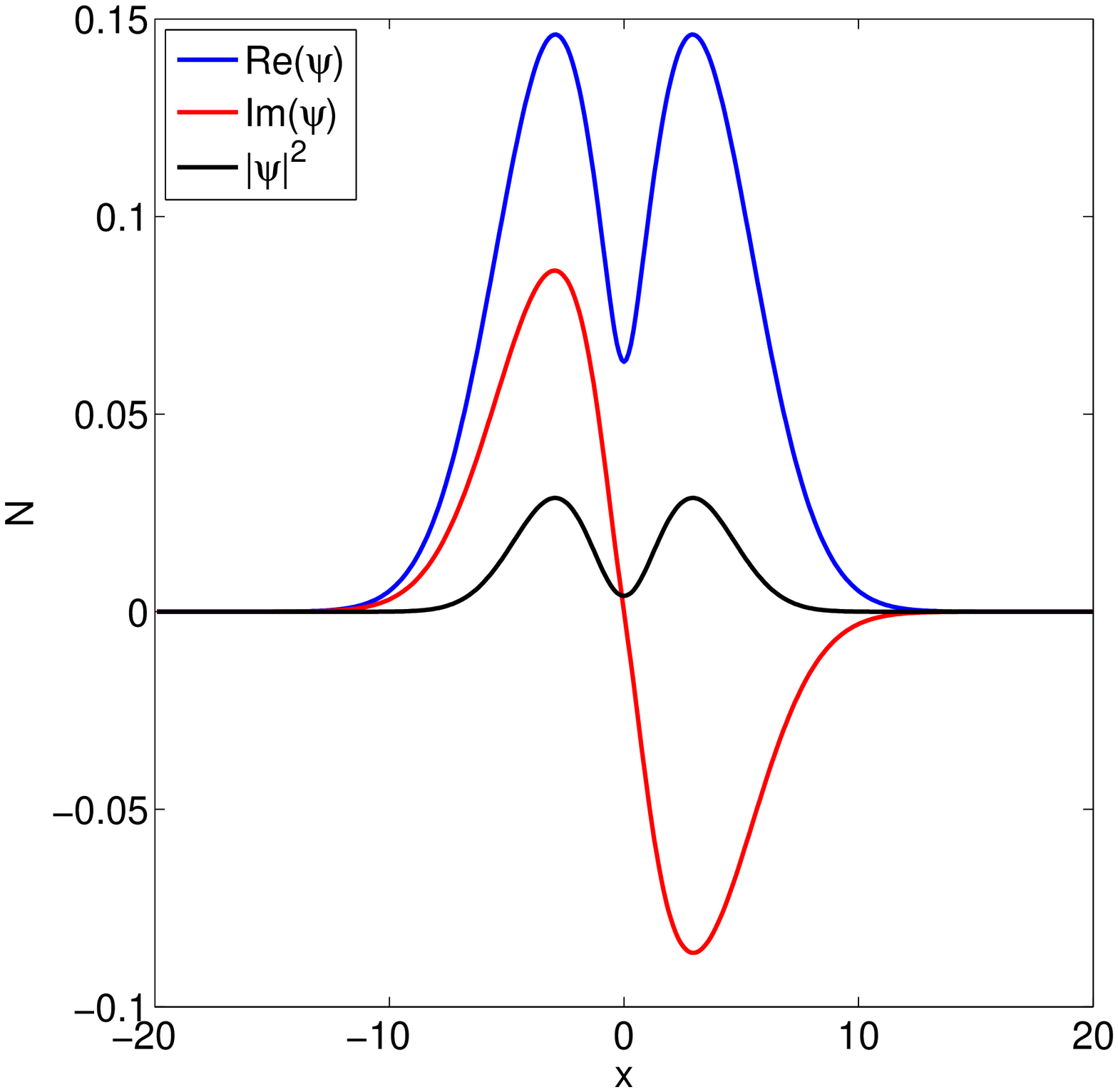}
\includegraphics[width=41mm,keepaspectratio]{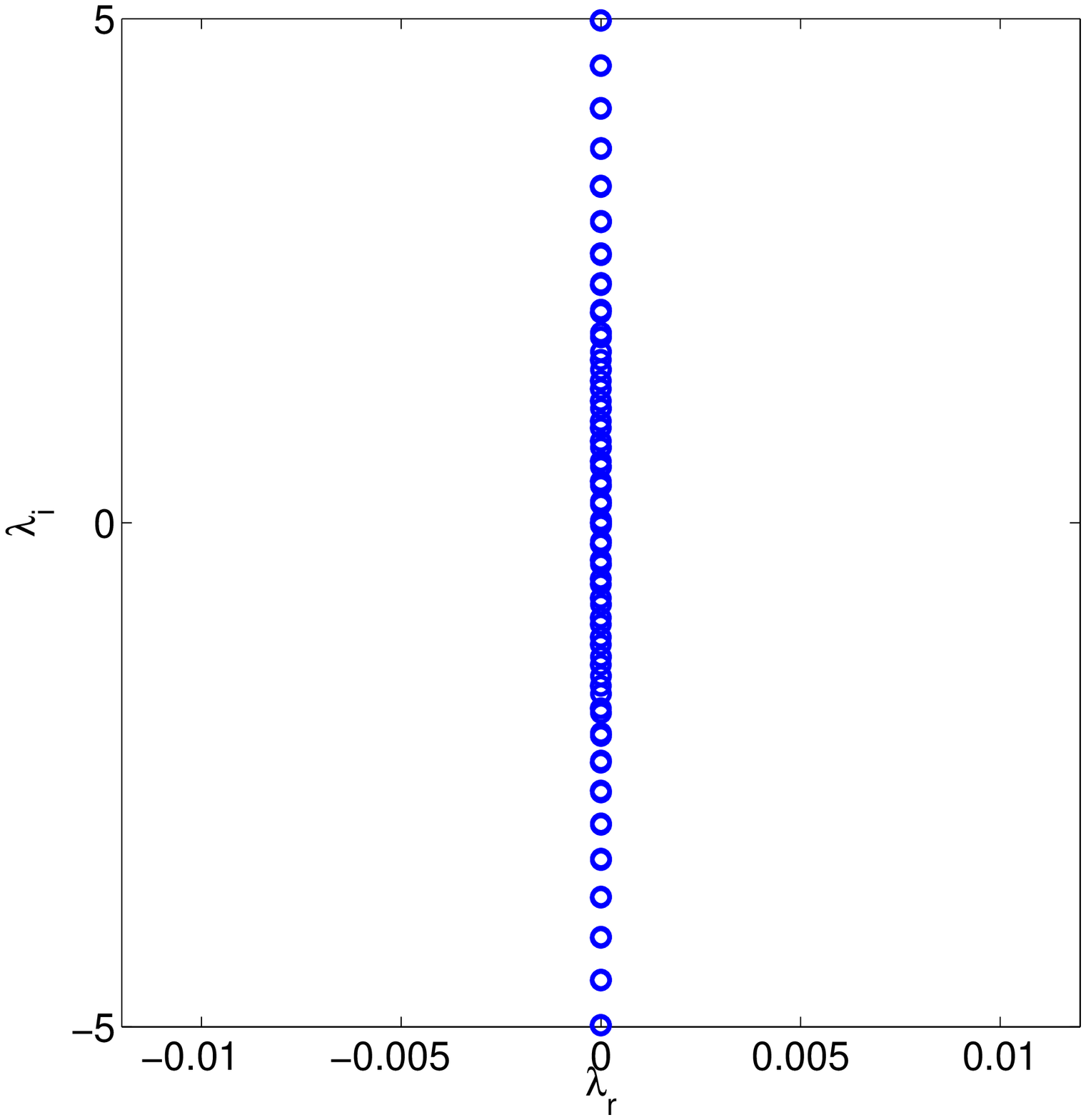}
\includegraphics[width=41mm,keepaspectratio]{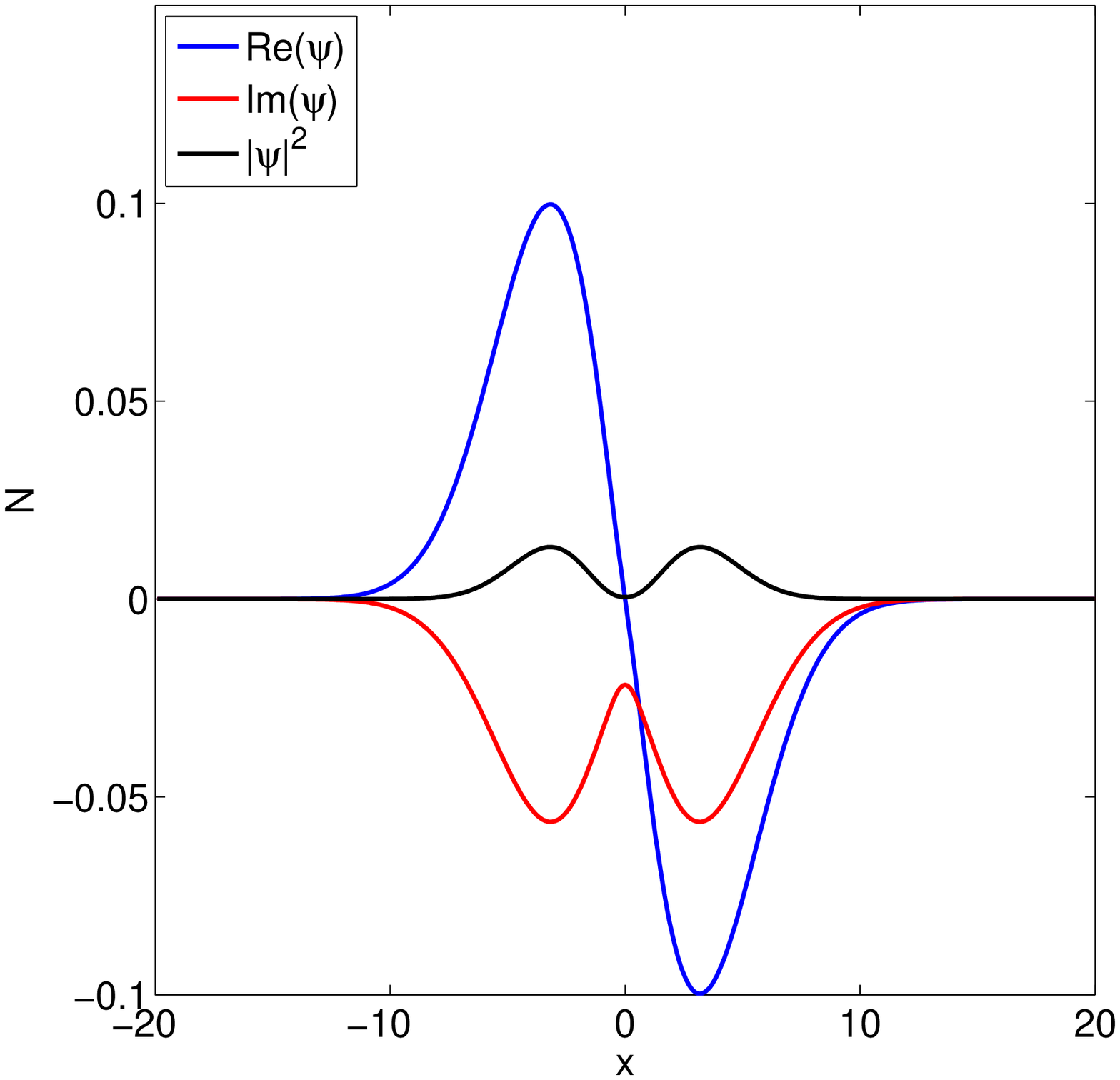}
\includegraphics[width=41mm,keepaspectratio]{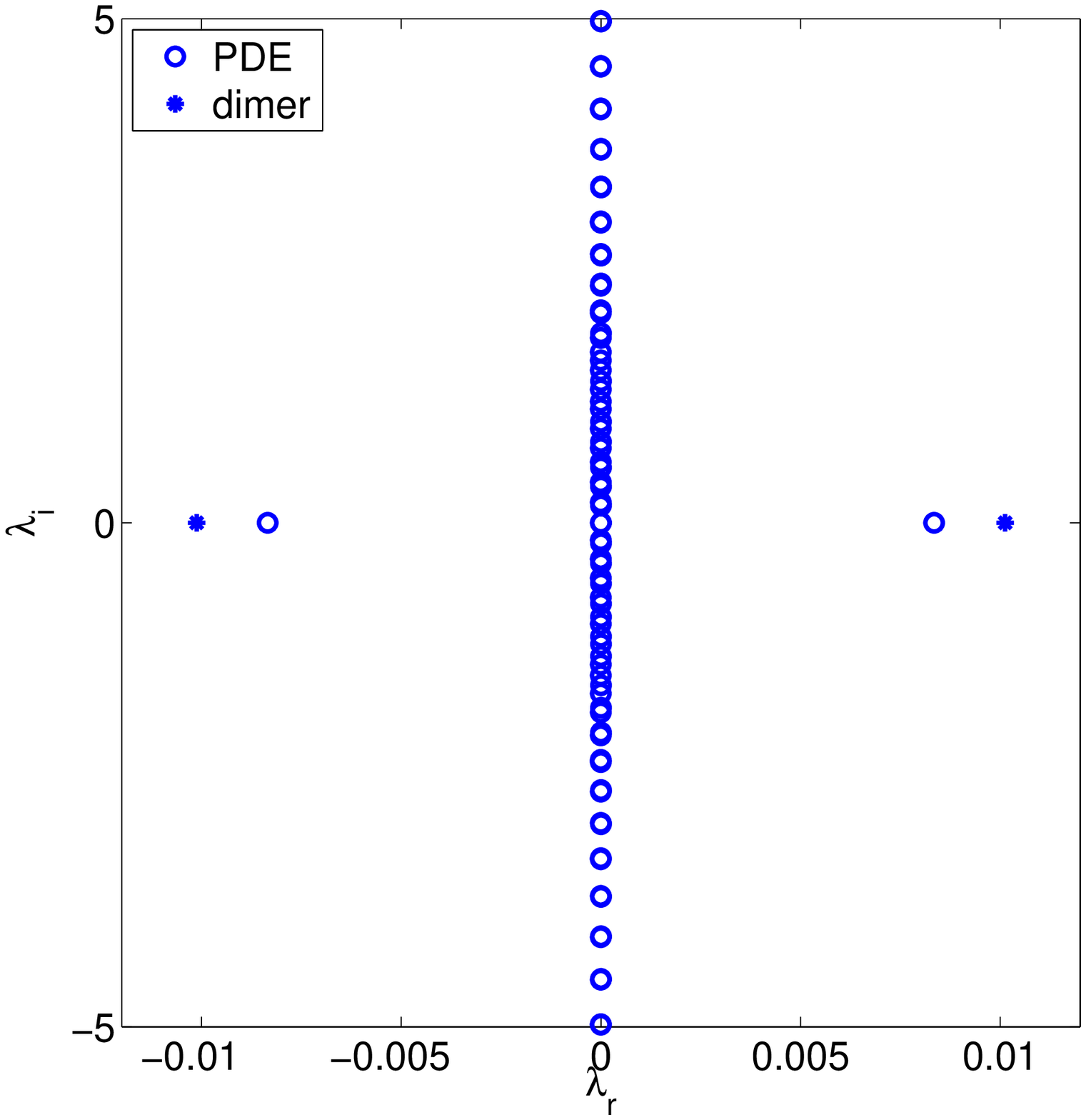}
\caption{(Color online) Real and imaginary parts of the states with real
(nonlinear) eigenvalues or equivalently real chemical potential. The left panel
presents the real and imaginary parts of the states and their square modulus,
while the right panel presents their corresponding eigenvalues of the
linearization. The top panel is for the symmetric state, while the bottom is for
the antisymmetric state. Both profiles are for the value of $\gamma=0.0100$ for
which the antisymmetric state is unstable. The two-mode prediction for the
relevant instability eigenvalue is indicated by stars.}
\label{fig2}
\end{figure}

There is a good agreement between the full PDE model and the DNLS dimer results 
despite the multiple approximations employed (from the original PDE to the
two-mode system and then from the two-mode system to the dimer). The main
features of the bifurcation diagram are as follows. For small values of $\gamma$
there are only two states with real chemical potential corresponding to the
symmetric and antisymmetric ones for $\gamma=0$. These states are the top and
the bottom one, respectively, coming from the limiting value $\gamma=0$. As we
approach the $\cP\cT$ phase transition (by increasing $\gamma$), the symmetric
branch develops an antisymmetric imaginary part (see top left panel of
Fig.~\ref{fig2}). Conversely, the antisymmetric branch develops a symmetric
imaginary part (see bottom left panel of Fig.~\ref{fig2}). At the critical point
the two waveforms are a $\pi/2$ rotation of one another; thus, they are
functionally equivalent in our U(1)-invariant setting. These branches collide
and disappear in the saddle-center bifurcation, which is the nonlinear analog of
the $\cP\cT$ symmetry-breaking phase transition. The critical point for this
transition is found to be at $0.01160\pm 0.00005$ from the PDE, while it is at
$0.01145$ in the $\cP\cT$-symmetric dimer picture.

\begin{figure}[tbp]
\includegraphics[width=41mm,keepaspectratio]{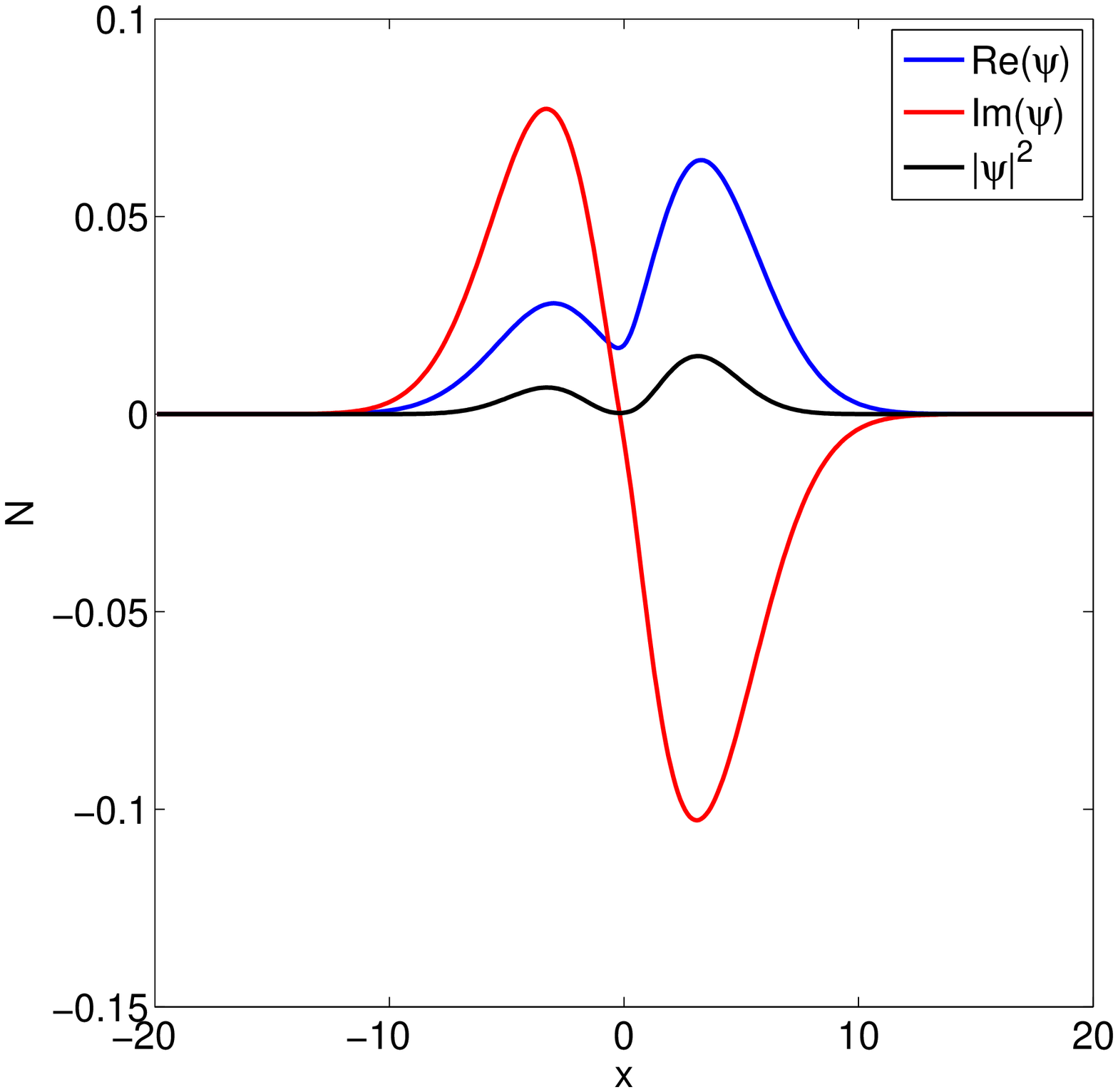}
\includegraphics[width=41mm,keepaspectratio]{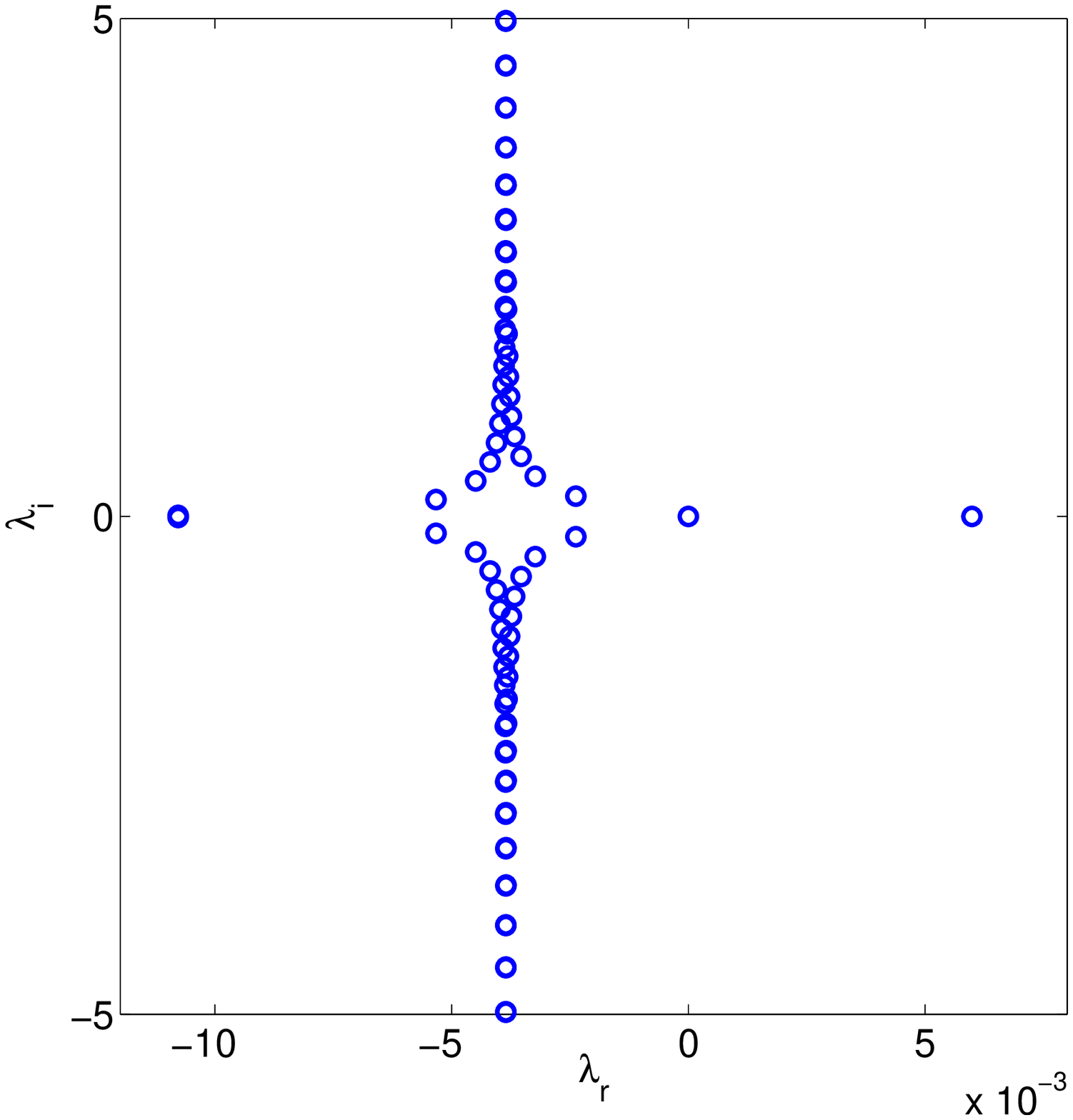}
\includegraphics[width=41mm,keepaspectratio]{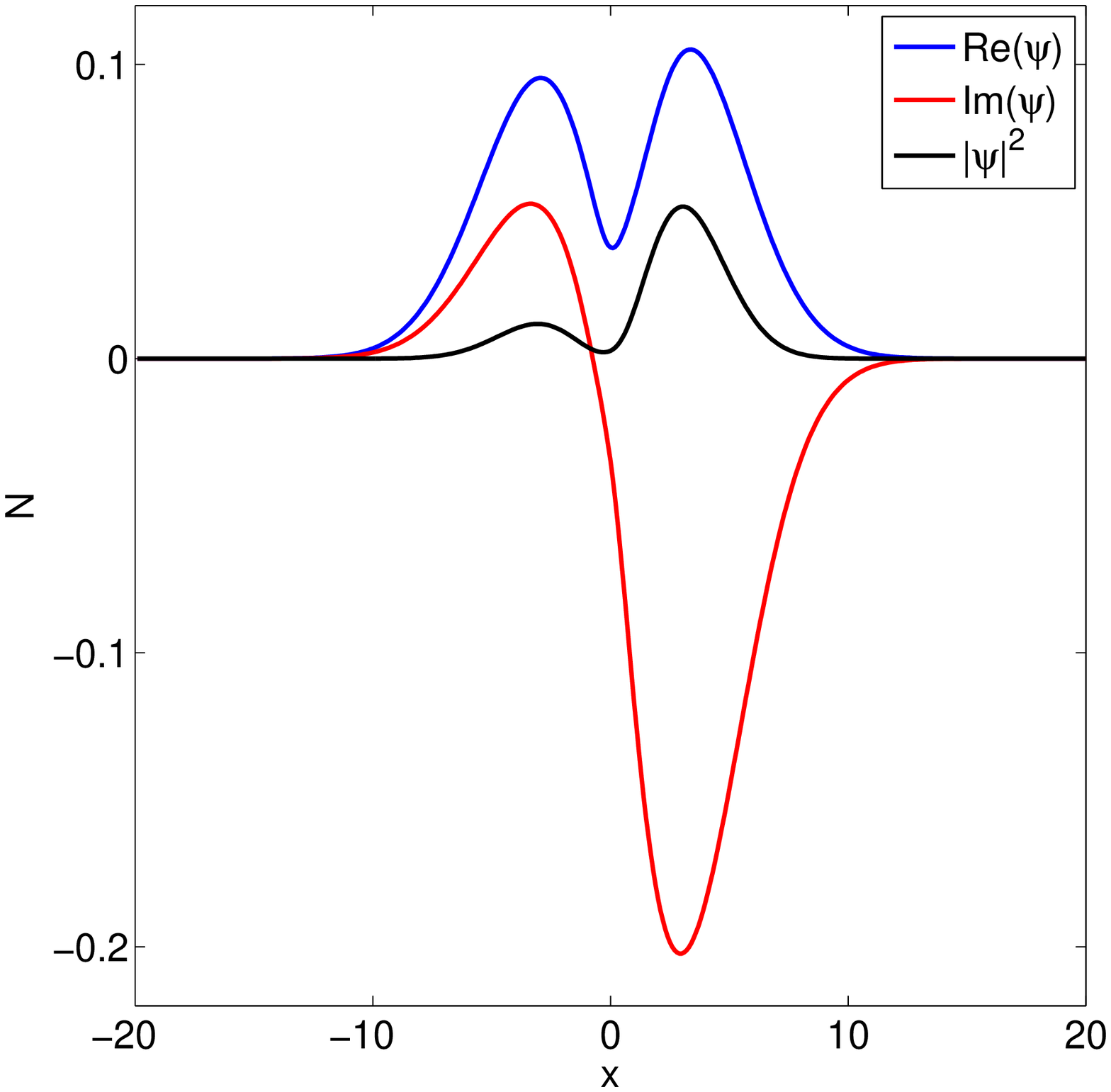}
\includegraphics[width=41mm,keepaspectratio]{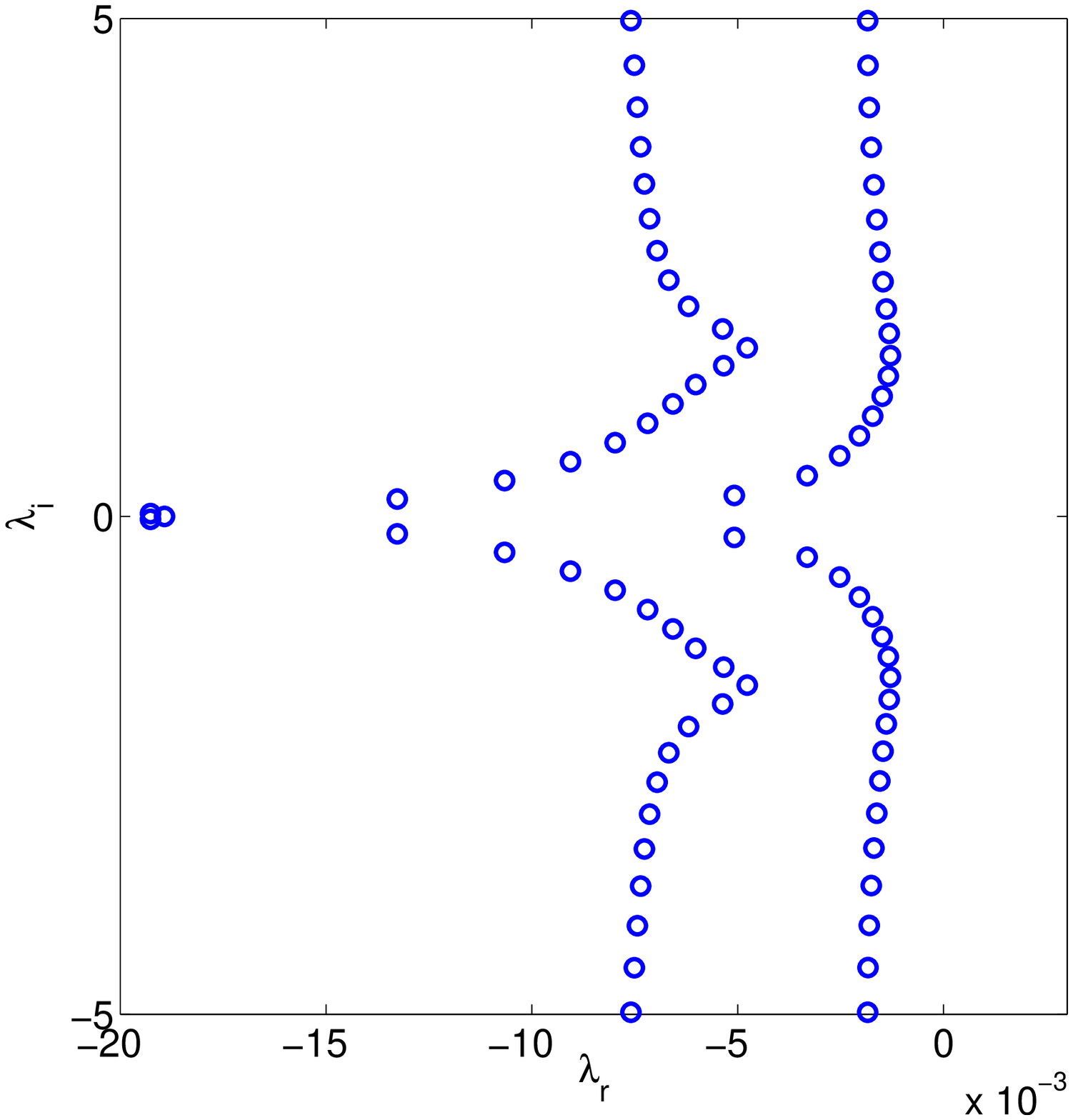}
\caption{(Color online) Same as Fig.~\ref{fig2}, but now the two states have an
imaginary part in their chemical potential. The top panel corresponds to the
profile (left) and the spectral plane of eigenvalues (right) for the ghost state
stemming from the bifurcation given again at $\gamma=0.0100$. The bottom panels,
which closely resemble to the top ones, are for the state stemming from the
analytic continuation computed for $\gamma=0.0150$.}
\label{fig4}
\end{figure}

Before this collision occurs, the antisymmetric state becomes unstable (turning
into the saddle of the saddle-center bifurcation) at the critical point $\gamma=
0.0083=\sqrt{k^2-E^2/4}$. The corresponding symmetry-breaking bifurcation for
the PDE occurs at $\gamma=0.00888$. The result of this symmetry breaking in a
purely nonlinear (non-$\cP\cT$ symmetric) setting would have been the
bifurcation of asymmetric states. Asymmetric states arise as well, but these are
the {\it ghost states} discussed above. These have been computed numerically and
are shown in the top left panel of Fig.~\ref{fig4}. A prototypical
characteristic of these states, shown in the bottom panel of Fig.~\ref{fig1}, is
the imaginary part $\mu_I$ of the chemical potential $\mu$ of these states,
which clearly demonstrates the pitchfork character of the bifurcation.

The parameter $\mu_I$ is self-consistently computed as follows. Recall that
stationary solutions of (\ref{e15}), including the ghost solutions of complex
$\mu$, are governed by
\begin{eqnarray}
\mu u={\cal L}u+iV_{\cP\cT}u+|u|^2u,
\label{e21}
\end{eqnarray}
while the conjugate equation reads
\begin{eqnarray}
\mu^*u^*={\cal L}u^*-iV_{\cP\cT}u^*+|u|^2 u^*.
\label{e22}
\end{eqnarray}
Then, multiplying (\ref{e21}) by $u^*$ and (\ref{e22}) by $u$, integrating, and
subtracting the second equation from the first, we obtain the self-consistency
condition for the imaginary part $\mu_I$ of the chemical potential:
\begin{eqnarray}
\mu_I=\frac{\int_{-\infty}^\infty dx\,V_{\cP\cT}|u|^2}{\int_{-\infty}^\infty dx
\,|u|^2}.
\label{e23}
\end{eqnarray}
This parameter corresponds to $E \sin(\phi_E)$ in the case of the dimer; that
is, there is a direct analogy between (\ref{e23}) with (\ref{e12}), which
becomes evident upon using the two-mode ansatz and the fact that $\gamma=\int dx
\,V_{\cP\cT}(x)u_L^2$. In this variable $\mu_I$ one can directly recognize the
pitchfork nature of the bifurcation at $\gamma=0.00888$ and the ghost nature of
these states. In fact, only one of these is expected to survive (the one with
positive $\mu_I$, which leads to growth and is denoted in bold), while the one
with negative $\mu_I$ decays and is not be observed in direct numerical
simulations. These symmetry-broken states are mirror images of one another, but
only one has the large amplitude at the ``right'' side (for $x>0$, where there
is indeed gain) (see the top panel of Fig.~\ref{fig4}). The other state has the
large amplitude at the ``wrong'' side (i.e., for $x<0$, where there is loss). 

The ghost states continue to exist past the critical point $\gamma=\pm k$ (as
they are not subject to a bifurcation at that critical point). However, the
other question is what becomes of the saddle and the center states (the former
antisymmetric and former symmetric one) past the critical point of the $\cP\cT$
phase transition. Following the suggestion of Ref.~\cite{R34} and the fact that
it is possible to require for both of the states with real $\mu$ (existing prior
to the $\cP\cT$ phase transition) that $u^*(x)=u(-x)$, we {\it enforce} this
condition for $\gamma>k$. This provides for the original model an {\it analytic
continuation} that is nonlocal in a highly nontrivial way; the steady-state
equations read
\begin{equation}
\mu u(x)={\cal L}u(x)+iV_{\cP\cT}(x)u(x)+u^2(x)u(-x),
\label{e24}
\end{equation}
and there is cross talk of the field value at $x$ and $-x$ within the nonlinear
term. These states {\it fall back} on the original states for $\gamma<k$, but
also provide their analytic continuation past this point. Interestingly, in this
case, a new inner product is used to compute the energy (atom number) $N$. Given
the substitution above, the energy now becomes $N=\int_{-\infty}^\infty dx\,u(x)
u(-x)$ (see also the review \cite{R2}). Also, from Eq.~(\ref{e23}) we obtain
\begin{equation}
\mu_I=\frac{\int_{-\infty}^\infty dx\,V_{\cP\cT}u(x)u(-x)}{\int_{-\infty}^\infty
dx\,u(x)u(-x)}.
\label{e25}
\end{equation}

The theoretical dimer predictions for the ghost branch and for the analytic
continuation branch are also shown in terms of $N$ as a function of $\gamma$ in
Fig.~\ref{fig1}. In the former case of the ghost states, the normalized (based
on the scaling transformation) norm is tantamount to $(A^2+B^2)/\eta_L\equiv(\mu
-\Omega)/\eta_L$, which is a horizontal line shown in Fig.~\ref{fig1}
bifurcating off of the theoretical point of $\gamma=0.0083$. On the other hand, 
for the latter case of the analytic continuation states, we have $N=\int dx\,u(x
)u(-x)=2c_Lc_R=2AB/\eta_L=2(\mu-\Omega)/\eta_L$. While there is a slight decay
of $N$ in the PDE for both states, the result is close to the theoretical
prediction of the dimer model. A similar weak dependence of a quantity similar
to $N$ on $\gamma$ was observed in the focusing case \cite{R34}.

\begin{figure}
\includegraphics[width=60mm,keepaspectratio]{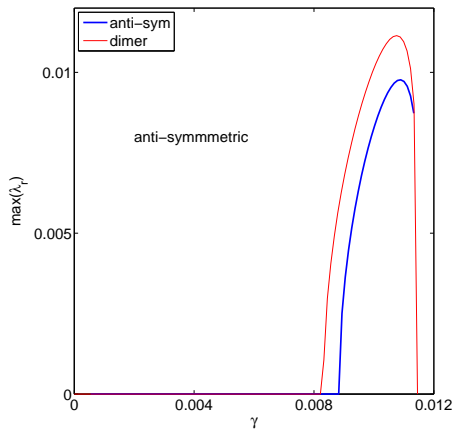}
\caption{(Color online) Maximum value of the real part of the eigenvalues as a
function of $\gamma$ for the antisymmetric state. The instability sets in at
$\gamma=0.00888$; for comparison, the corresponding eigenvalue is 
also evaluated from the dimer reduction.}
\label{fig6}
\end{figure}

Detailed examples of all branches are shown in Figs.~\ref{fig2} - \ref{fig4}. As
described above, the symmetric branch shows a symmetric real part and an
antisymmetric imaginary part, while the opposite occurs for the antisymmetric
branch. Both branches are stable; up to the critical point no eigenvalues with
positive real part arise. At that point, in the case of the antisymmetric
branch, a pair of imaginary eigenvalues moves to the real axis (as shown in the
bottom right panel of Fig.~\ref{fig2}). The variation of the maximum real part
of the eigenvalues of the antisymmetric branch as a function of $\gamma$ is
shown in Fig.~\ref{fig6}. In both of the above panels the comparison with the
analytical approximation given by $\lambda=\pm2i\sqrt{2(k^2-\gamma^2)-E\sqrt{k^2
-\gamma^2}}$ is provided, which gives a measure of the agreement with the
two-mode discrete picture.

The eigenvalues of the ghost state that emerges from the supercritical pitchfork
bifurcation are shown in the top right panel of Fig.~\ref{fig4}. Here, the
solution is unstable too, due to a positive real eigenvalue. Importantly, one
also sees here a shift of neutrally stable eigenvalues to the left-half plane,
where they lead to decaying excitations. This spectral picture is valid for the
branch of positive $\mu_I$. It should nevertheless be recalled that this
stability analysis is not of direct use as these states are {\it not}
stationary, and the existence of positive $\mu_I$ leads to an amplitude that
grows in time. Similar conclusions can be drawn for the states resulting from
the analytic continuation of the model. These states are presented in the bottom
panel of Fig.~\ref{fig4}; these states structurally resemble the ghost states,
particularly in the profile of the square moduli. Although the model used to
compute the latter is different and involves the nonlinearity $u^2(x)u(-x)$,
both states have nonzero (positive in this case) $\mu_I$, and both terminate at
the same critical point as $\gamma$ is increased in the two-mode analysis. This
structural proximity of the profiles of the top and the bottom panel of
Fig.~\ref{fig4} can be interpreted as a byproduct of the predicted proximity
for large $\gamma$ in the two-mode picture.

\begin{figure}[tbp]
\includegraphics[width=42mm,keepaspectratio]{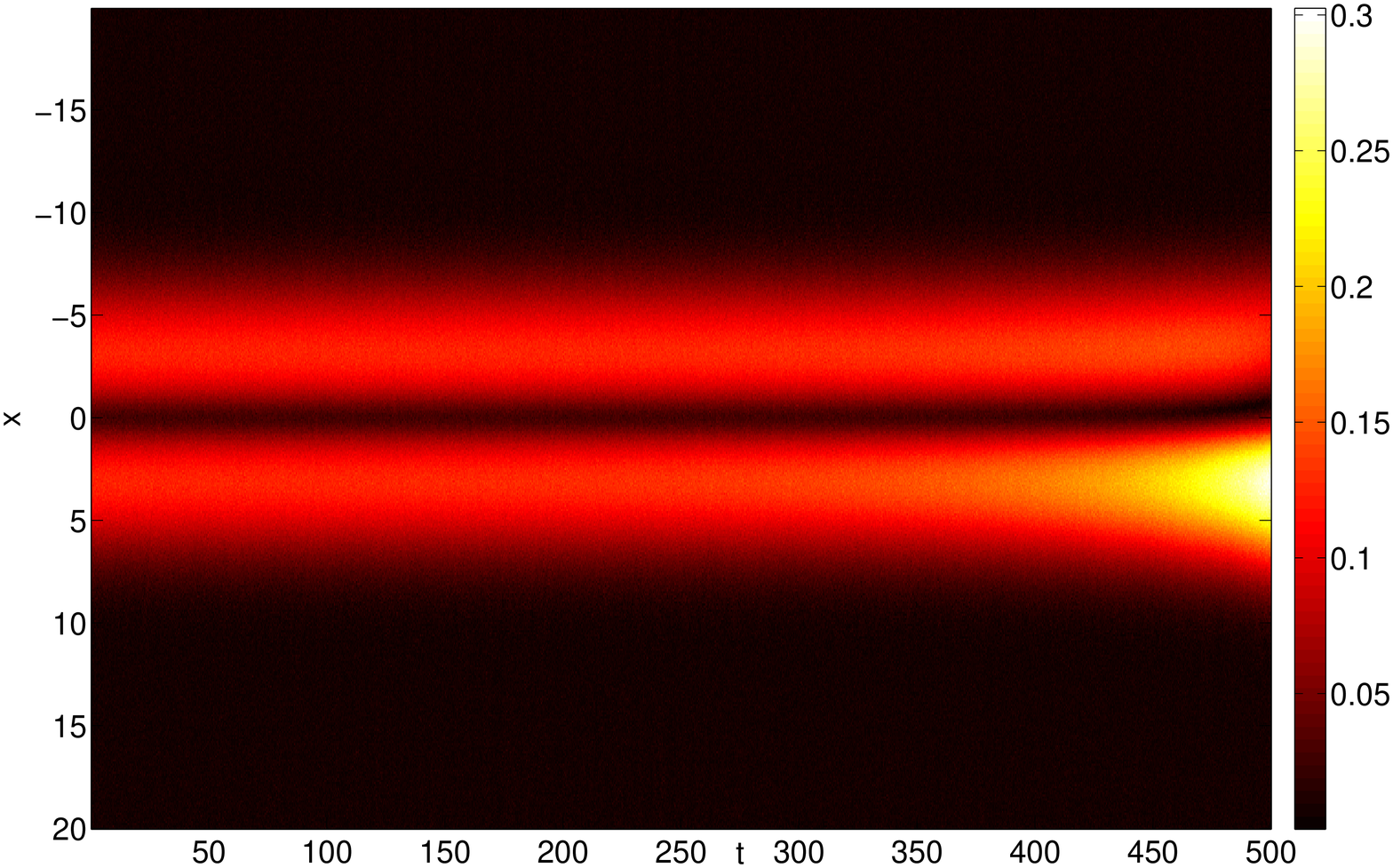}
\includegraphics[width=42mm,keepaspectratio]{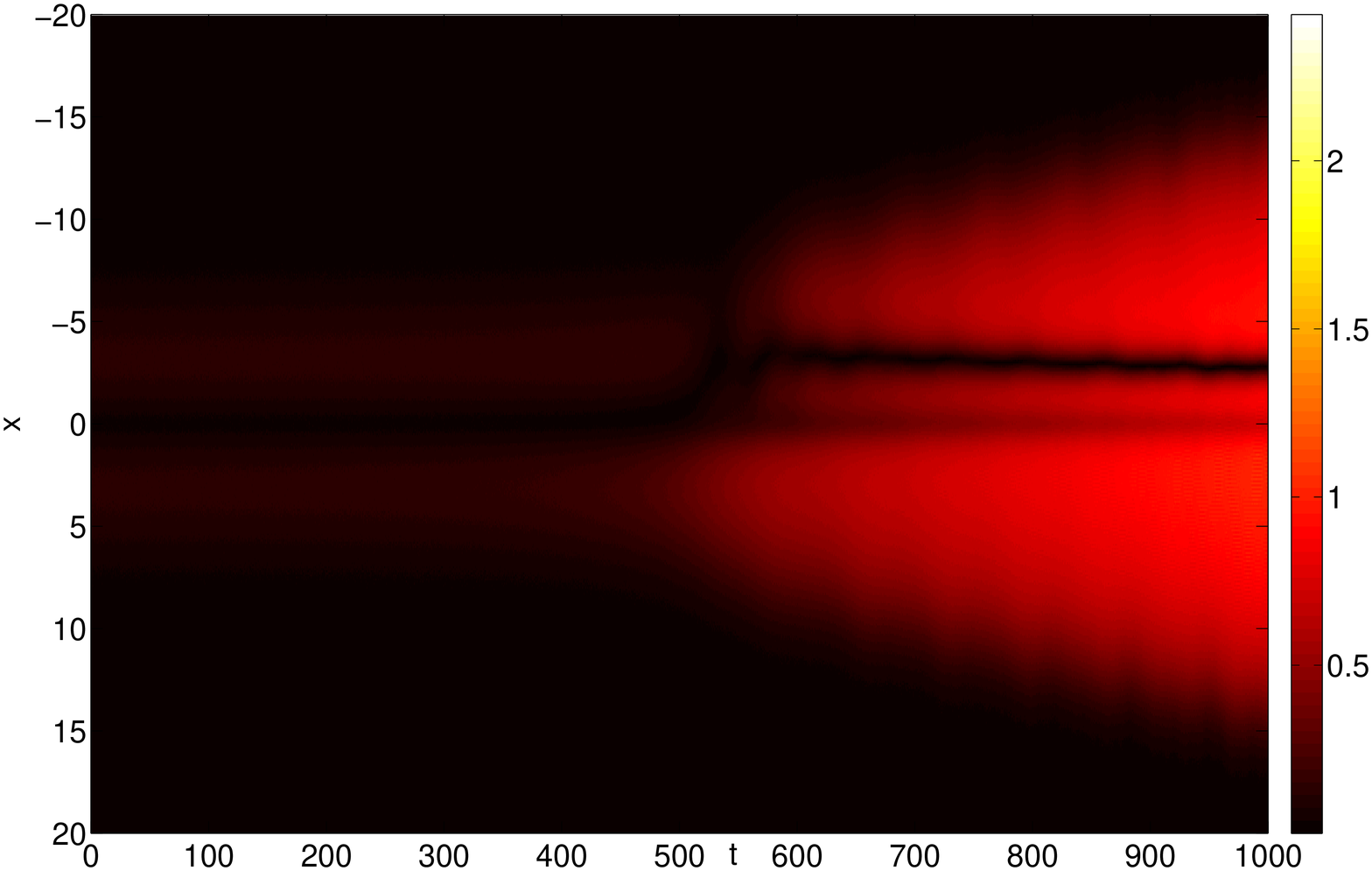}
\includegraphics[width=42mm,keepaspectratio]{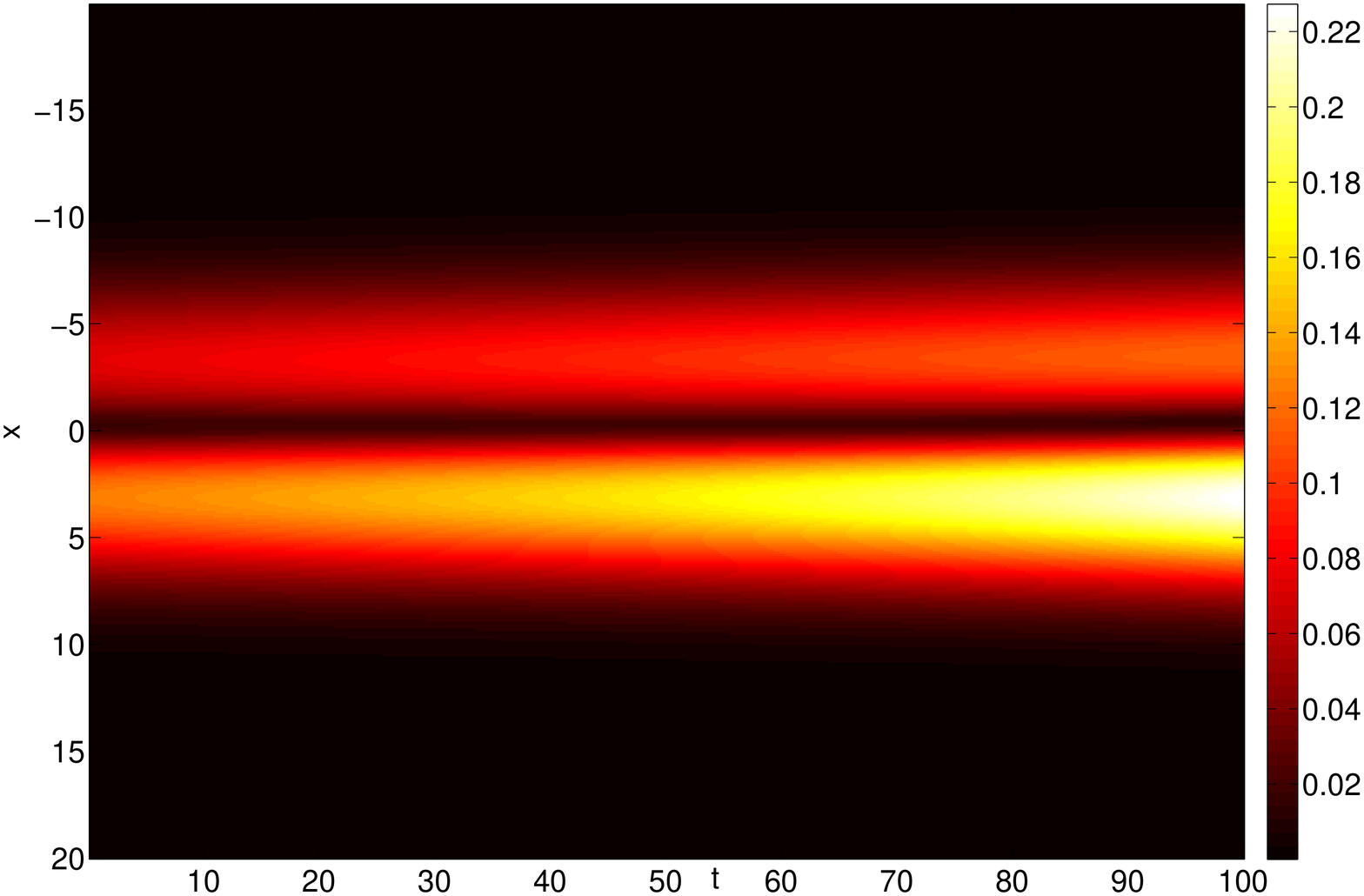}
\includegraphics[width=42mm,keepaspectratio]{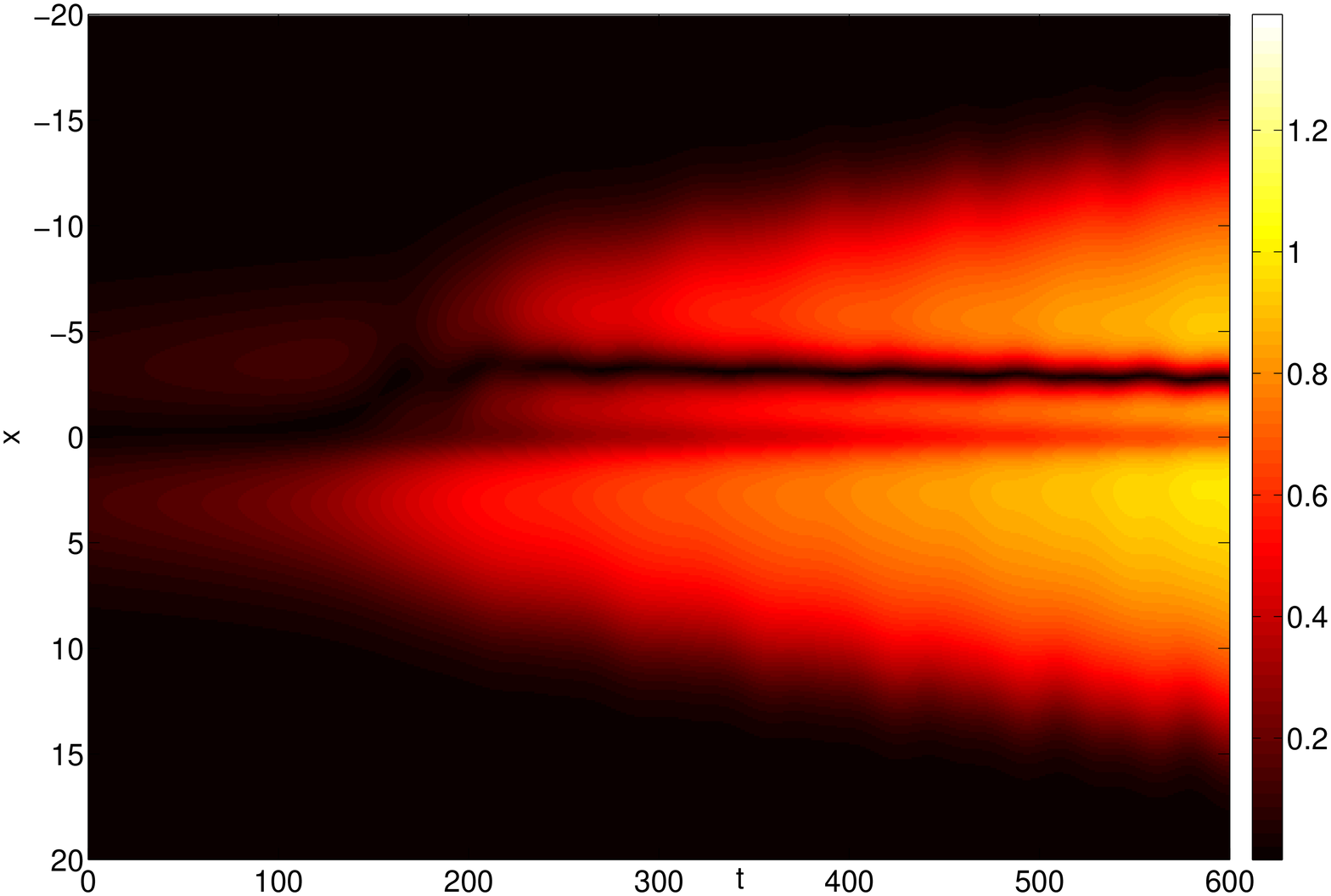}
\caption{(Color online) Contour plots showing the space-time evolution of the
density $|u(x,t)|^2$ (for $\gamma=0.01090$). The top panels show the evolution
of the antisymmetric state, illustrating its dynamical destabilization, and the
bottom panels show the evolution of the ghost state. (Note the different time
scales between the panels.) The left panels show early stages of the evolution
(initial destabilization between $t=0$ and $t=400$ for the antisymmetric state
and then growth from $t=400$ to $t=500$). The initial stage only shows the
growth in the case of the ghost state enabled by $\mu_I>0$. The right panels
show the late stages of the evolution where dynamics, intractable within the
two-mode model, develop. This evolution involves a dark-soliton state nucleated
within the left well, similar to what was observed in Refs.~\cite{R40,R41}.}
\label{fig7}
\end{figure}

Finally, we studied the dynamics of the states by means of direct numerical
simulations. Prior to the critical point, we examined the dynamical evolution of
the instability of the antisymmetric state, as shown in Fig.~\ref{fig7} for
$\gamma=0.01090$. The top left panel illustrates a short time scale over which
the mode becomes unstable. Note that the evolution of the antisymmetric mode
between $t=400$ and $t=500$ closely emulates the growth of the ghost mode
between $t=0$ and $t=100$. Its instability appears to follow the growth of the
corresponding ghost mode, which is shown in the bottom left panel of
Fig.~\ref{fig7}. For longer time scales, the ghost mode itself becomes unstable
and approaches another ghost state of the type that has been illustrated in
Refs.~\cite{R40,R41} (see the bottom right panel of Fig.~\ref{fig7}). The latter
state involves a dark soliton that has migrated to one of the wells of the
double-well potential, accompanied by a growth of the amplitude and the width of
the solution. Unfortunately, this type of state is intractable within the
two-mode picture (as the latter does not allow for the possibility of intra-well
dynamics). In the same way, the antisymmetric state, after transiting through
its corresponding (for the same $\gamma$) ghost state, follows the instability
and fate of the ghost state by forming for longer times the same tilted
dark-soliton dynamical structure (see the top right panel of Fig.~\ref{fig7}).

\begin{figure}[h]
\includegraphics[width=42mm,keepaspectratio]{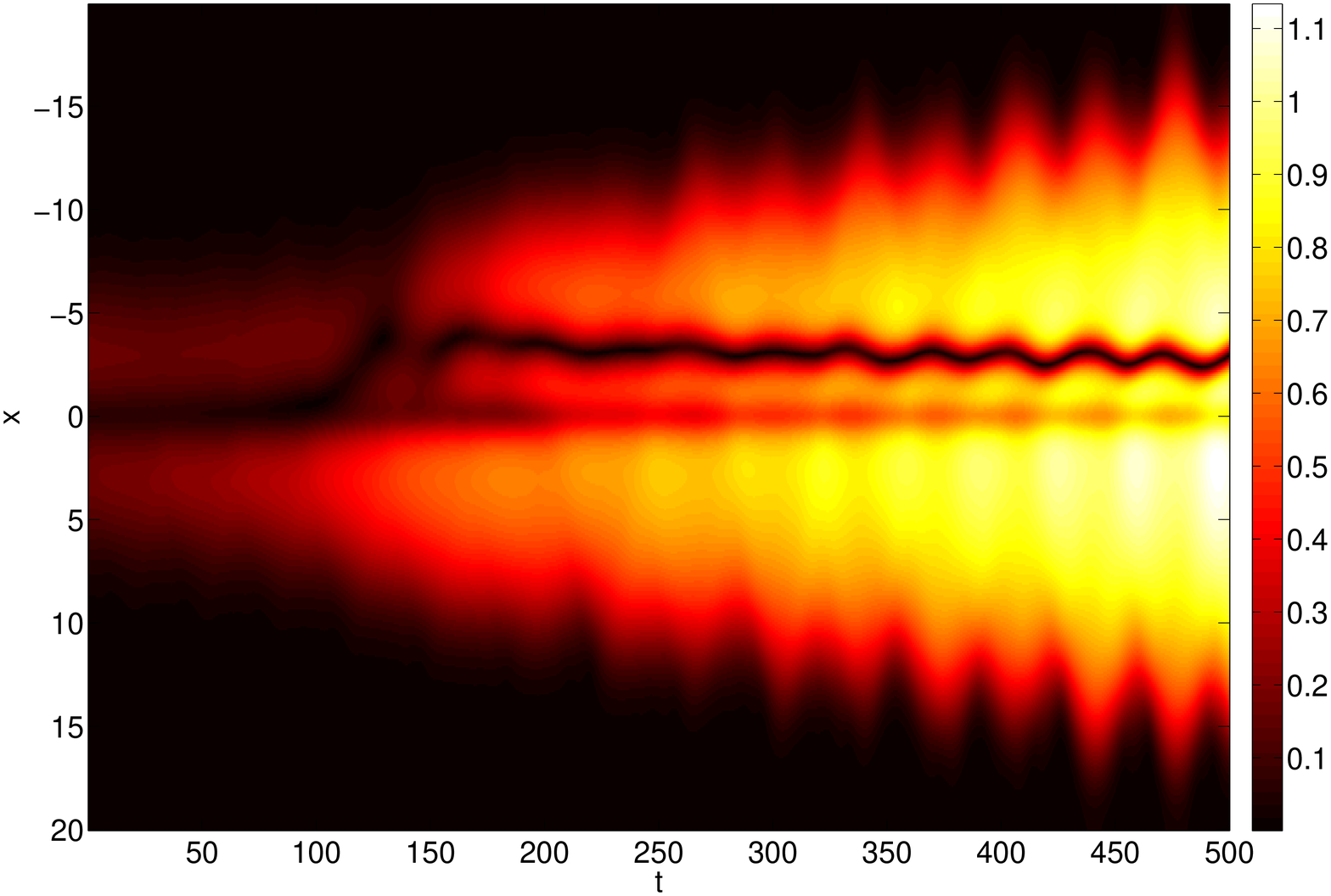}
\includegraphics[width=42mm,keepaspectratio]{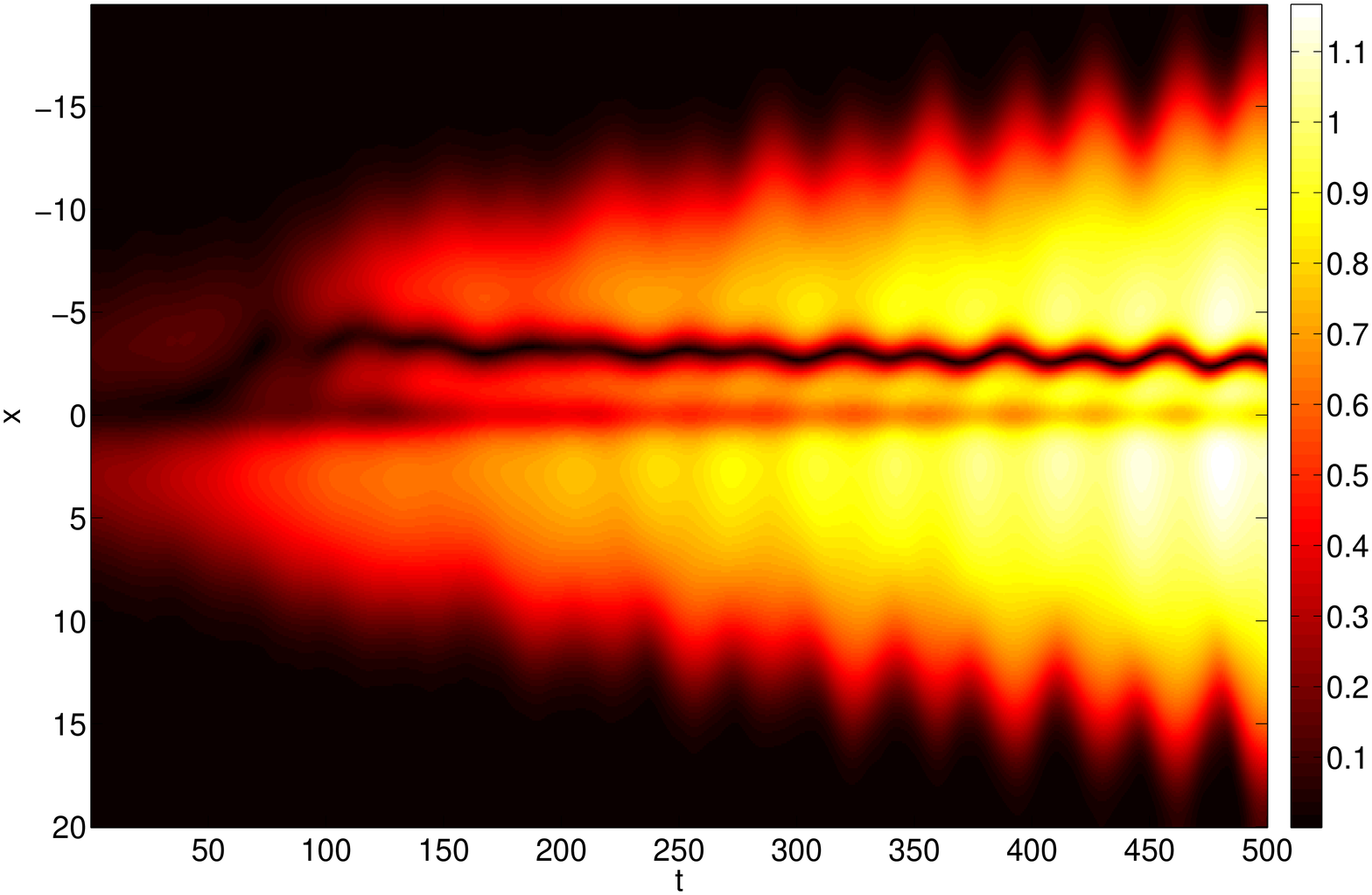}
\caption{(Color online) Left panel: Space-time evolution of an exact solution of
the symmetric branch obtained for $\gamma=0.0100$, when initialized in
(\ref{e15}) for $\gamma=0.0150$, i.e., past the critical point of the
transition, where the symmetric branch no longer exists. Right panel: Evolution
of the exact ghost state obtained for $\gamma=0.0150$.}
\label{fig9}
\end{figure}

Having considered the dynamical instability prior to the $\cP\cT$ phase
transition critical point, we now turn to the dynamics of various states past
the latter point of $\gamma=0.01160$. Above the critical point (and, in
particular, for $\gamma=0.0150$) we performed the evolution of three different
initial waveforms. The first initialization (whose evolution is shown in the
left panel of Fig.~\ref{fig9}) used the symmetric solution as obtained for a
smaller value of $\gamma$, namely for $\gamma=0.0100$, to examine the outcome of
the ``standard'' waveforms past the $\cP\cT$-phase-transition point which leads
to the termination of their existence. This dynamics illustrated a phenomenology
similar to that shown earlier in Fig.~\ref{fig7}. The initial growth stage was
finally succeeded by the formation of a robust dark soliton on the left well of
the potential. A similar evolution shown in the right panel of the figure was
observed for the case of the ghost state with $\gamma=0.0150$. We also performed
a similar computation initialized with the result of the analytic continuation
and a similar integration result was obtained (not shown). We conclude that past
the critical point of the $\cP\cT$ phase transition, the dynamics is typically
attracted to dark soliton ghost states, which reside within the lossy well (and
can thus not be captured by our two-mode picture).

\section{Conclusions and outlook}

In this paper we have used analytical and numerical methods to revisit the
problem of a $\cP\cT$-symmetric double-well potential. We used a particular form
for the symmetric real part and the antisymmetric ($\cP\cT$-symmetric) imaginary
part of the potential, but we indicated that the specific choice of the
potential is inconsequential. In fact, given the double-well structure and a
suitable choice of parameters (such as the chemical potential, i.e., the
strength of the nonlinearity), one can reduce the problem to a two-mode regime.
In that setting a reduction can be made explicit (and its
assumptions/transformations can be suitably formulated) from the original PDE
problem (the nonlinear Schr\"odinger model with a $\cP\cT$-symmetric double-well
potential) to the simpler and analytically tractable dimer setting.

In the latter, we have shown that all characteristics of the system become
transparent and we have identified the symmetric and antisymmetric states and
their nonlinear continuation. We have quantified the pitchfork bifurcation,
formerly leading to asymmetric states due to nonlinearity, and presently leading
to ghost states due to $\cP\cT$-symmetry. We have explicitly obtained and
monitored the daughter states of this bifurcation and their role in the
dynamics, even though they are not exact solutions. We computed the $\cP\cT$
phase transition (or simply the saddle-center bifurcation) of the two
fundamental branches. Finally, the analytic continuation of the system past this
critical point, with its unusual ``nonlocal'' (at least at the PDE level) cross
talk of $x$ and $-x$ was formulated. In all of the above, the analytics
performed in the framework of the two-mode approximation remain invariably a
reasonable approximation to the full original system and the numerical
computations therein.

The above analysis gives a fundamental picture for the combined existence and
interplay of nonlinearity, effective discreteness (through the double-well
potential) and $\cP\cT$-symmetry. However, there are numerous directions for
interesting generalizations of this work. One of these would be to consider the
more complicated ``oligomer'' configurations of Ref.~\cite{R28}; namely, the
trimers and quadrimers (for the latter, see also Ref.~\cite{R42} and even the
plaquette two-dimensonal building blocks of Ref.~\cite{R43}). Another aspect
that has emerged both here and in the context of Ref.~\cite{R41} is to
investigate the dynamics of the asymmetric states in which a dark soliton
emerges and localizes within the lossy region. These are ghost states of the
full PDE model; as we indicated, they cannot be captured by the two-mode
approximation. These ghost states deserve a systematic investigation and
classification. Generalizing to higher-dimensional settings, and introducing
topologically charged states such as vortices~\cite{R40,R41} would also be worth
exploring. These topics are presently under study and results will be reported
in future publications.

PGK gratefully acknowledges the support of the U.S. National Science Foundation
under grants DMS-0806762 and CMMI-1000337, as well as the Alexander von Humboldt
Foundation, the Alexander S. Onassis Public Benefit Foundation and the
Binational Science Foundation. DJF was partially supported by the Special
Account for Research Grants of the University of Athens. CMB is supported by the
U.S. Department of Energy and the U.K. Leverhulme Foundation.

{\bf Note Added}: During the finalization stage of the present work, the authors
became aware of the recently posted manuscript Ref.~\cite{R43} that addresses
(albeit from a slightly different perspective) the static problem associated
with the discussion of section II.B, namely the $\cP\cT$-symmetric dimer.

\end{document}